\documentclass[12pt]{article}
\usepackage{amssymb}
\usepackage{graphicx}
\usepackage{indentfirst}
\usepackage{cite}

\def\gammac{c\!\!\!/}
\def\gammau{u\!\!\!/}

\begin{document}

\title{Early thermalization of quark-gluon matter  \\
with the elastic scattering of $ggq$ and $gg\bar q$}
\author{Xiao-Ming Xu$^1$, Zhen-Yu Shen$^1$, Zhi-Cheng Ye$^2$, and
Wei-Jie Xu$^3$}
\date{}
\maketitle \vspace{-1cm}
\centerline{$^1$Department of Physics, Shanghai University,
Baoshan, Shanghai 200444, China}
\centerline{$^2$Department of Mathematics, Shanghai University,
Baoshan, Shanghai 200444, China}
\centerline{$^3$Department of Transportation Engineering, Shanghai Maritime
University,}
\centerline{Pudong New District, Shanghai 201306, China}
%----------------------------------------------------------------
\begin{abstract}
Elastic gluon-gluon-quark (gluon-gluon-antiquark) scattering is studied in
perturbative QCD with 123 Feynman diagrams at the tree level. Individually
squared amplitudes and interference terms of the Feynman diagrams are derived.
With the elastic gluon-gluon-quark scattering and the elastic 
gluon-gluon-antiquark scattering transport equations are established. In the
thermalization process of initially created quark-gluon matter, this matter is
governed by elastic 2-to-2 scattering and elastic 3-to-3 scattering.
Solutions of the transport equations show that initially created quark-gluon
matter takes early thermalization, i.e., thermal states are established
rapidly. Different thermalization times of gluon matter and quark matter are
obtained.
\end{abstract}
\noindent
PACS: 24.85.+p; 12.38.Mh; 25.75.Nq

\noindent
Keywords: Quark-gluon matter; Elastic gluon-gluon-quark scattering; Transport
equation; Thermalization

\newpage

\leftline{\bf I. INTRODUCTION}
\vspace{0.5cm}

Quark-gluon plasma is a thermal state and has a temperature, but quark-gluon
matter initially produced in Au-Au collisions at high energies of the
Relativistic Heavy Ion Collider (RHIC)
is not in a thermal state and does not have a temperature.
Therefore, how and when initially produced quark-gluon matter establishes a
thermal state are key questions in evolution from initially produced
quark-gluon matter to quark-gluon plasma, and the evolution is a thermalization
process. Hydrodynamic calculations \cite{PHHH,TLS}
in explaining the experimental elliptic flow coefficient of hadrons 
\cite{STAR} have revealed that the thermalization time is less than 1 
fm/$c$, i.e., the thermalization is rapid \cite{PHENIX}. Initially produced 
quark-gluon matter has anisotropic distributions of quarks and gluons. The
distribution function in the direction of the incoming gold beam is much larger
than that in the direction perpendicular to the beam direction \cite{XMX}.
Massless gluons and quarks produced in initial Au-Au collisions move at the
velocity of light. Intuitively, such fast gluons and quarks can not become
randomly moving, i.e., a thermal state cannot be established easily. But the
early thermalization is against the intuition. Therefore, the early
thermalization is a new phenomenon and must be understood! 
The way to understand the early
thermalization is to answer the two questions: how initially produced
quark-gluon matter establishes a thermal state and when initially produced
quark-gluon matter establishes a thermal state.

Let us notice that the gluon number density of quark-gluon matter initially
produced in central Au-Au collisions at RHIC energies can be very high
\cite{CMN}. A high number density must have its consequences. Indeed, the
early thermalization is closely related to the high number density 
\cite{XMX,XSCZ}. Initially produced gluon matter can be
simulated by HIJING \cite{WG} and in the collision region corresponds to
a gluon number density of 19.4 ${\rm fm}^{-3}$. At an interaction range of 0.62
fm, which acts as the radius of a unit volume of sphere, gluon-gluon
scattering and gluon-gluon-gluon scattering have occurrence probabilities
of 0.3 and 0.2, respectively \cite{XMCW}. This implies substantial occurrence
of the gluon-gluon-gluon scattering, and the 3-gluon scattering and the 2-gluon
scattering lead to the early thermalization of initially produced gluon matter
\cite{XSCZ}. The 2-gluon scattering alone can not cause 
the early thermalization, and
the early thermalization is understood as an effect of many-body scattering.

In Ref. \cite{XX} we have studied the thermalization of initially produced
quark-gluon matter with elastic 2-to-2 scattering which happens no matter what
gluon and quark number densities are \cite{shuryak,wong,SM,XG,BGLMV}
and elastic 3-to-3 scattering which includes the elastic scattering of 
gluon-gluon-gluon, quark-quark-quark, antiquark-antiquark-antiquark, 
quark-quark-antiquark, quark-antiquark-antiquark, gluon-quark-quark, 
gluon-antiquark-antiquark, and gluon-quark-antiquark. 
Since it is hard to derive squared amplitudes for elastic
gluon-gluon-quark scattering and elastic gluon-gluon-antiquark scattering,
we can include the two types of elastic scattering only in the present work. 
Feynman diagrams at the tree level for the elastic gluon-gluon-quark
scattering are introduced in Sec. II. A Feynman diagram has a squared
amplitude. Two diagrams have two individually squared amplitudes and one
interference term. The individually squared amplitudes and the interference
term are derived in perturbative QCD in Sec. III. The sum of the individually
squared amplitudes and the interference terms of all Feynman diagrams
is the squared amplitude for the elastic gluon-gluon-quark scattering. The
squared amplitude for the elastic gluon-gluon-antiquark scattering is identical
with the one for the elastic gluon-gluon-quark scattering.
Including the squared amplitudes for the two types of elastic scattering,
transport equations for gluons, quarks and antiquarks are given in
Sec. IV. Numerical solutions of the transport equations and discussions are
presented in Sec. V. A summary is in the last section.

\vspace{0.5cm}
\leftline{\bf II. ELASTIC GLUON-GLUON-QUARK SCATTERING}
\vspace{0.5cm}

The elastic gluon-gluon-quark scattering brings about 123 Feynman diagrams at
the tree level. The 123 diagrams are divided into four classes. The first class
consists of 24 diagrams that do not contain triple-gluon vertices and
four-gluon vertices; the second class 36 diagrams of which each contains one
triple-gluon vertex and no one four-gluon vertex; the third class 38 diagrams
of which each contains either two triple-gluon vertices or one four-gluon
vertex; the fourth class 25 diagrams of which each contains either three
triple-gluon vertices or one four-gluon vertex plus one triple-gluon vertex.
Let wiggly lines and solid lines stand for gluons and quarks, respectively.
Twenty-one Feynman diagrams are displayed in Figs. 1-6. The other one hundred
and two diagrams are derived from the diagrams in Figs. 1-6 as follows.

\vspace{0.5cm}
\centerline{\bf A. Feynman diagrams in the first class}
\vspace{0.5cm}

The Feynman diagrams in the first class involve the absorption processes of
initial gluons and the emission processes of final gluons. Every diagram in the
first class has four gluon-quark vertices.
The Feynman diagram in Fig. 1 indicates that the left initial gluon is absorbed
first by the quark, the right initial gluon is then absorbed, the
left final gluon is emitted first from the quark, and the right final gluon is
then emitted. This is a permutation of the absorption processes of the two
initial gluons and the emission processes of the two final gluons. If the right
final gluon is emitted first and is followed by the absorption of the right
initial gluon, and the left final gluon is the next emitted and is followed by
the absorption of the left initial gluon, another permutation of the absorption
and emission processes is obtained. The corresponding Feynman diagram is
derived from the diagram in Fig. 1 by the exchange of the vertex of the left
initial gluon and the quark and the vertex of the right final gluon and the
quark. From this example we infer that the diagram ${\rm C}_{1245}$ leads to 23
new diagrams by exchanging the four gluon-quark
vertices. Therefore, the 24 diagrams in the first class correspond to 24
permutations of the absorption amd emission processes.

\vspace{0.5cm}
\centerline{\bf B. Feynman diagrams in the second class}
\vspace{0.5cm}

Any of the Feynman diagrams in the second class involves one triple-gluon
vertex and no one four-gluon vertex. The triple-gluon vertex in a diagram
in Fig. 2 relates to the two initial gluons,
one initial gluon and one final gluon, or the two final gluons. We may derive
33 diagrams from the three diagrams.

The diagram ${\rm C}_{(12)45}$ in Fig. 2 indicates that the two initial gluons
annihilate into one gluon that is absorbed by the quark, and the quark then
radiates two gluons. The diagram has three gluon-quark vertices and
leads to five new diagrams with permutations of the three vertices. Hence, we 
have 6 Feynman diagrams each of which has a triple-gluon
vertex that possesses the two initial gluons. 

The diagram ${\rm C}_{(14)25}$ means that one initial gluon breaks into one 
final gluon and a virtual gluon that is absorbed by the quark, and, 
furthermore,
the quark absorbs another initial gluon and then radiates another final gluon.
Now we take into account the following five cases: (1) the gluon that breaks
into two gluons may be the left or right initial gluon; (2) the final gluon
that comes from the broken gluon is either the left or right final
gluon; (3) the initial gluon absorbed by the quark may be the left or right
initial gluon; (4) the final gluon radiated is either the left or right final
gluon; (5) the three gluon-quark vertices may take six permutations. Operating
along with the five cases, we get 24 diagrams which include the diagram 
${\rm C}_{(14)25}$. Each of the 24 diagrams has a triple-gluon vertex that 
possesses one initial gluon and one final gluon.

The diagram ${\rm C}_{21(45)}$ indicates that the quark absorbs the two initial
gluons first and then radiates a gluon which breaks into
the two final gluons. The quark line has three gluon-quark vertices. Different
permutations of the three vertices lead to five new diagrams from the diagram
${\rm C}_{21(45)}$. Hence, we have 6 Feynman diagrams each of which has a
triple-gluon vertex that possesses the two final gluons.

\vspace{0.5cm}
\centerline{\bf C. Feynman diagrams in the third class}
\vspace{0.5cm}

Thirty diagrams in the third class involve two triple-gluon vertices
individually, and the other eight diagrams one four-gluon vertex. Six of the
thirty diagrams are shown in Fig. 3, and two of the eight diagrams in Fig. 4.
The diagram ${\rm C}_{(25)(14)}$ or ${\rm C}_{(12)(45)}$ uses up the external
gluons in the two triple-gluon vertices.
In each of the other diagrams there is one external gluon coupled to the quark.

The six diagrams in Fig. 3 are characteristic of having two triple-gluon
vertices and two gluon-quark vertices. The permutation of the two gluon-quark
vertices generates six new diagrams. More diagrams can be generated from
the exchange of the two initial gluons in the diagrams ${\rm C}_{(142)5}$,
${\rm C}_{2(145)}$, and ${\rm C}_{2(451)}$ and/or the
exchange of the two final gluons in the diagrams ${\rm C}_{(124)5}$,
${\rm C}_{(142)5}$, ${\rm C}_{2(145)}$, and ${\rm C}_{(25)(14)}$. We can thus
derive 24 diagrams from the six diagrams in Fig. 3.

The two diagrams in Fig. 4 are characteristic of one four-gluon vertex and two
gluon-quark vertices. The two gluon-quark vertices may be exchanged. The two
final gluons in the diagram ${\rm C}_{(124\sim )5}$ can be exchanged, so can
the two initial gluons in the diagram ${\rm C}_{2(145\sim )}$. The four
exchanges lead to six new diagrams from the two diagrams in Fig. 4.

\vspace{0.5cm}
\centerline{\bf D. Feynman diagrams in the fourth class}
\vspace{0.5cm}

Fifteen diagrams in the fourth class relate to three triple-gluon vertices
individually, and the other ten diagrams one triple-gluon vertex plus one
four-gluon vertex. Four of the fifteen diagrams are shown in Fig. 5, and five
of the ten diagrams in Fig. 6. In every diagram the quark line is only coupled
to a virtual gluon.

In addition to the three triple-gluon vertices every diagram in Fig. 5 has
three virtual gluons, but only one virtual gluon is connected to the quark. 
In the diagram ${\rm C}_{(12)(45)2}$ the virtual gluon is also connected to the
right initial gluon. Of course, the virtual gluon can be connected to
the other three external gluons, and three new diagrams are thus generated. We
cannot derive a diagram from the diagram ${\rm C}_{(12)(45){\rm M}}$,
but can derive one more diagram from the diagram ${\rm C}_{(14)(25){\rm M}}$ by
the exchange of the two final gluons. One virtual gluon in the
diagram ${\rm C}_{14-}$ connects the
quark and the left final gluon. It may also connect the quark and the
other three external gluons. The two final gluons may be exchanged.
Then, we can derive 7 diagrams from the diagram ${\rm C}_{14-}$. In total, we
have derived 11 diagrams from the four diagrams in Fig. 5.

Every diagram in Fig. 6 has one triple-gluon vertex and one four-gluon vertex.
If the two initial gluons are coupled in a vertex and the two final gluons
in another vertex as shown in the diagrams
${\rm C}_{12\sim}$ and ${\rm C}_{45\sim}$, no more diagrams can be derived.
If three external gluons are coupled in a vertex as
shown in the diagrams ${\rm C}_{(12)45\sim}$ and ${\rm C}_{12(45)\sim}$,
the exchange of the two external gluons mounted in different vertices
leads to two new diagrams. If  every vertex is mounted with an initial gluon
and a final gluon as shown in the diagram ${\rm C}_{25\sim}$, the exchange of
the two initial gluons and/or the exchange of
the two final gluons generate three new diagrams.
In total, we can derive 5 diagrams from the five diagrams in Fig. 6.

\vspace{0.5cm}
\leftline{\bf III. SQUARED AMPLITUDE FOR ELASTIC GLUON-GLUON-QUARK}
\leftline{\bf SCATTERING}
\vspace{0.5cm}

Amplitudes of the Feynman diagrams of the elastic gluon-gluon-quark
scattering $g(p_1)+g(p_2)+q(p_3) \to 
g(p_4)+g(p_5)+q(p_6)$ with four-momenta $p_i=(E_i,\vec{p}_i)(i=1,\cdots,6)$
are written in accordance with the
Feynman rules in perturbative QCD \cite{field,CS}. The amplitudes contain
the polarization four-vectors $\epsilon_\mu (\lambda_i)(i=1,2,4,5)$ of
external gluons
and linear combinations of gluon momenta due to the triple-gluon vertices.
A diagram's squared amplitude or an interference term of two diagrams
has a structure with one of the following forty-one types:
\begin{displaymath}
{\rm tr} (\gammau_1 \gammau_2 \gammau_3 \gammau_4 \gammau_5
\gammau_6 \gammau_7 \gammau_8 \gammau_9 \gammau_{10} \gammau_{11} \gammau_{12}
\gammau_{13} \gammau_{14} \gammau_{15} \gammau_{16}),
\end{displaymath}
\begin{displaymath}
\epsilon^* (\lambda_{k_1}) \cdot \epsilon^* (\lambda_{k_2})
{\rm tr} (\gammau_1 \gammau_2 \gammau_3 \gammau_4 \gammau_5
\gammau_6 \gammau_7 \gammau_8 \gammau_9 \gammau_{10} \gammau_{11} \gammau_{12}
\gammau_{13} \gammau_{14}),
\end{displaymath}
\begin{displaymath}
u_{15} \cdot \epsilon^* (\lambda_{k_1})
{\rm tr} (\gammau_1 \gammau_2 \gammau_3 \gammau_4 \gammau_5
\gammau_6 \gammau_7 \gammau_8 \gammau_9 \gammau_{10} \gammau_{11} \gammau_{12}
\gammau_{13} \gammau_{14}),
\end{displaymath}
\begin{displaymath}
\epsilon (\lambda_{k_1}) \cdot \epsilon (\lambda_{k_2})
\epsilon^* (\lambda_{k_3}) \cdot \epsilon^* (\lambda_{k_4})
{\rm tr} (\gammau_1 \gammau_2 \gammau_3 \gammau_4 \gammau_5 \gammau_6
\gammau_7 \gammau_8 \gammau_9 \gammau_{10} \gammau_{11} \gammau_{12}),
\end{displaymath}
\begin{displaymath}
u_{13} \cdot u_{14} \epsilon^* (\lambda_{k_1}) \cdot \epsilon^* (\lambda_{k_2})
{\rm tr} (\gammau_1 \gammau_2 \gammau_3 \gammau_4 \gammau_5 \gammau_6 
\gammau_7 \gammau_8 \gammau_9 \gammau_{10} \gammau_{11} \gammau_{12}),
\end{displaymath}
\begin{displaymath}
u_{13} \cdot \epsilon (\lambda_{k_1}) u_{14} \cdot \epsilon^* (\lambda_{k_2})
{\rm tr} (\gammau_1 \gammau_2 \gammau_3 \gammau_4 \gammau_5 \gammau_6 
\gammau_7 \gammau_8 \gammau_9 \gammau_{10} \gammau_{11} \gammau_{12}),
\end{displaymath}
\begin{displaymath}
\epsilon (\lambda_{k_1}) \cdot \epsilon (\lambda_{k_2})
u_{13} \cdot \epsilon^* (\lambda_{k_3})
{\rm tr} (\gammau_1 \gammau_2 \gammau_3 \gammau_4 \gammau_5 \gammau_6 
\gammau_7 \gammau_8 \gammau_9 \gammau_{10} \gammau_{11} \gammau_{12}),
\end{displaymath}
\begin{displaymath}
\epsilon (\lambda_{k_1}) \cdot \epsilon (\lambda_{k_2})
\epsilon^* (\lambda_{k_3}) \cdot \epsilon^* (\lambda_{k_4})
\epsilon^* (\lambda_{k_5}) \cdot \epsilon^* (\lambda_{k_6})
{\rm tr} (\gammau_1 \gammau_2 \gammau_3 \gammau_4 \gammau_5
\gammau_6 \gammau_7 \gammau_8 \gammau_9 \gammau_{10}),
\end{displaymath}
\begin{displaymath}
u_{11} \cdot u_{12} \epsilon^* (\lambda_{k_1}) \cdot \epsilon^* (\lambda_{k_2})
\epsilon^* (\lambda_{k_3}) \cdot \epsilon^* (\lambda_{k_4})
{\rm tr} (\gammau_1 \gammau_2 \gammau_3 \gammau_4 \gammau_5
\gammau_6 \gammau_7 \gammau_8 \gammau_9 \gammau_{10}),
\end{displaymath}
\begin{displaymath}
u_{11} \cdot \epsilon^* (\lambda_{k_1}) u_{12} \cdot \epsilon^* (\lambda_{k_2})
u_{13} \cdot \epsilon^* (\lambda_{k_3})
{\rm tr} (\gammau_1 \gammau_2 \gammau_3 \gammau_4 \gammau_5
\gammau_6 \gammau_7 \gammau_8 \gammau_9 \gammau_{10}),
\end{displaymath}
\begin{displaymath}
\epsilon^* (\lambda_{k_1}) \cdot \epsilon^* (\lambda_{k_2})
u_{11} \cdot \epsilon^* (\lambda_{k_3}) u_{12} \cdot \epsilon^* (\lambda_{k_4})
{\rm tr} (\gammau_1 \gammau_2 \gammau_3 \gammau_4 \gammau_5
\gammau_6 \gammau_7 \gammau_8 \gammau_9 \gammau_{10}),
\end{displaymath}
\begin{displaymath}
u_{11} \cdot u_{12} \epsilon^* (\lambda_{k_1}) \cdot \epsilon^* (\lambda_{k_2})
u_{13} \cdot \epsilon^* (\lambda_{k_3})
{\rm tr} (\gammau_1 \gammau_2 \gammau_3 \gammau_4 \gammau_5
\gammau_6 \gammau_7 \gammau_8 \gammau_9 \gammau_{10}),
\end{displaymath}
\begin{displaymath}
\epsilon (\lambda_{k_1}) \cdot \epsilon (\lambda_{k_2})
\epsilon^* (\lambda_{k_3}) \cdot \epsilon^* (\lambda_{k_4})
u_{11} \cdot \epsilon^* (\lambda_{k_5})
{\rm tr} (\gammau_1 \gammau_2 \gammau_3 \gammau_4 \gammau_5
\gammau_6 \gammau_7 \gammau_8 \gammau_9 \gammau_{10}),
\end{displaymath}
\begin{displaymath}
\epsilon (\lambda_{k_1}) \cdot \epsilon (\lambda_{k_2})
\epsilon (\lambda_{k_3}) \cdot \epsilon (\lambda_{k_4})
\epsilon^* (\lambda_{k_5}) \cdot \epsilon^* (\lambda_{k_6})
\epsilon^* (\lambda_{k_7}) \cdot \epsilon^* (\lambda_{k_8})
{\rm tr} (\gammau_1 \gammau_2 \gammau_3 \gammau_4 \gammau_5
\gammau_6 \gammau_7 \gammau_8),
\end{displaymath}
\begin{displaymath}
u_{9} \cdot u_{10} u_{11} \cdot u_{12}
\epsilon (\lambda_{k_1}) \cdot \epsilon (\lambda_{k_2})
\epsilon^* (\lambda_{k_3}) \cdot \epsilon^* (\lambda_{k_4})
{\rm tr} (\gammau_1 \gammau_2 \gammau_3 \gammau_4 \gammau_5
\gammau_6 \gammau_7 \gammau_8),
\end{displaymath}
\begin{displaymath}
u_{9} \cdot u_{10} \epsilon (\lambda_{k_1}) \cdot \epsilon (\lambda_{k_2})
\epsilon^* (\lambda_{k_3}) \cdot \epsilon^* (\lambda_{k_4})
\epsilon^* (\lambda_{k_5}) \cdot \epsilon^* (\lambda_{k_6})
{\rm tr} (\gammau_1 \gammau_2 \gammau_3 \gammau_4 \gammau_5
\gammau_6 \gammau_7 \gammau_8),
\end{displaymath}
\begin{displaymath}
u_{9} \cdot \epsilon (\lambda_{k_1}) u_{10} \cdot \epsilon (\lambda_{k_2})
u_{11} \cdot \epsilon^* (\lambda_{k_3}) u_{12} \cdot \epsilon^* (\lambda_{k_4})
{\rm tr} (\gammau_1 \gammau_2 \gammau_3 \gammau_4 \gammau_5
\gammau_6 \gammau_7 \gammau_8),
\end{displaymath}
\begin{displaymath}
\epsilon (\lambda_{k_1}) \cdot \epsilon (\lambda_{k_2})
u_{9} \cdot \epsilon (\lambda_{k_3}) u_{10} \cdot \epsilon^* (\lambda_{k_4})
u_{11} \cdot \epsilon^* (\lambda_{k_5})
{\rm tr} (\gammau_1 \gammau_2 \gammau_3 \gammau_4 \gammau_5
\gammau_6 \gammau_7 \gammau_8),
\end{displaymath}
\begin{displaymath}
u_{9} \cdot u_{10} \epsilon^* (\lambda_{k_1}) \cdot \epsilon^* (\lambda_{k_2})
u_{11} \cdot \epsilon (\lambda_{k_3}) u_{12} \cdot \epsilon (\lambda_{k_4})
{\rm tr} (\gammau_1 \gammau_2 \gammau_3 \gammau_4 \gammau_5
\gammau_6 \gammau_7 \gammau_8),
\end{displaymath}
\begin{displaymath}
\epsilon (\lambda_{k_1}) \cdot \epsilon (\lambda_{k_2})
\epsilon^* (\lambda_{k_3}) \cdot \epsilon^* (\lambda_{k_4})
u_{9} \cdot \epsilon (\lambda_{k_5}) u_{10} \cdot \epsilon^* (\lambda_{k_6})
{\rm tr} (\gammau_1 \gammau_2 \gammau_3 \gammau_4 \gammau_5
\gammau_6 \gammau_7 \gammau_8),
\end{displaymath}
\begin{displaymath}
u_{9} \cdot u_{10} \epsilon (\lambda_{k_1}) \cdot \epsilon (\lambda_{k_2})
\epsilon^* (\lambda_{k_3}) \cdot \epsilon^* (\lambda_{k_4})
u_{11} \cdot \epsilon (\lambda_{k_5})
{\rm tr} (\gammau_1 \gammau_2 \gammau_3 \gammau_4 \gammau_5
\gammau_6 \gammau_7 \gammau_8),
\end{displaymath}
\begin{displaymath}
\epsilon (\lambda_{k_1}) \cdot \epsilon (\lambda_{k_2})
\epsilon^* (\lambda_{k_3}) \cdot \epsilon^* (\lambda_{k_4})
\epsilon^* (\lambda_{k_5}) \cdot \epsilon^* (\lambda_{k_6})
u_9 \cdot \epsilon (\lambda_{k_7})
{\rm tr} (\gammau_1 \gammau_2 \gammau_3 \gammau_4 \gammau_5
\gammau_6 \gammau_7 \gammau_8),
\end{displaymath}
\begin{displaymath}
u_7 \cdot u_8 u_9 \cdot u_{10}
\epsilon (\lambda_{k_1}) \cdot \epsilon (\lambda_{k_2})
\epsilon^* (\lambda_{k_3}) \cdot \epsilon^* (\lambda_{k_4})
\epsilon^* (\lambda_{k_5}) \cdot \epsilon^* (\lambda_{k_6})
{\rm tr} (\gammau_1 \gammau_2 \gammau_3 \gammau_4 \gammau_5 \gammau_6),
\end{displaymath}
\begin{displaymath}
u_7 \cdot u_8 \epsilon (\lambda_{k_1}) \cdot \epsilon (\lambda_{k_2})
\epsilon (\lambda_{k_3}) \cdot \epsilon (\lambda_{k_4})
\epsilon^* (\lambda_{k_5}) \cdot \epsilon^* (\lambda_{k_6})
\epsilon^* (\lambda_{k_7}) \cdot \epsilon^* (\lambda_{k_8})
{\rm tr} (\gammau_1 \gammau_2 \gammau_3 \gammau_4 \gammau_5 \gammau_6),
\end{displaymath}
\begin{displaymath}
u_7 \cdot \epsilon (\lambda_{k_1}) u_8 \cdot \epsilon (\lambda_{k_2})
u_9 \cdot \epsilon^* (\lambda_{k_3}) u_{10} \cdot \epsilon^* (\lambda_{k_4})
u_{11} \cdot \epsilon^* (\lambda_{k_5})
{\rm tr} (\gammau_1 \gammau_2 \gammau_3 \gammau_4 \gammau_5 \gammau_6),
\end{displaymath}
\begin{displaymath}
\epsilon (\lambda_{k_1}) \cdot \epsilon (\lambda_{k_2})
u_7 \cdot \epsilon (\lambda_{k_3}) u_8 \cdot \epsilon^* (\lambda_{k_4})
u_9 \cdot \epsilon^* (\lambda_{k_5}) u_{10} \cdot \epsilon^* (\lambda_{k_6})
{\rm tr} (\gammau_1 \gammau_2 \gammau_3 \gammau_4 \gammau_5 \gammau_6),
\end{displaymath}
\begin{displaymath}
u_7 \cdot u_8 \epsilon^* (\lambda_{k_1}) \cdot \epsilon^* (\lambda_{k_2})
u_9 \cdot \epsilon (\lambda_{k_3}) u_{10} \cdot \epsilon (\lambda_{k_4})
u_{11} \cdot \epsilon^* (\lambda_{k_5})
{\rm tr} (\gammau_1 \gammau_2 \gammau_3 \gammau_4 \gammau_5 \gammau_6),
\end{displaymath}
\begin{displaymath}
\epsilon (\lambda_{k_1}) \cdot \epsilon (\lambda_{k_2})
\epsilon^* (\lambda_{k_3}) \cdot \epsilon^* (\lambda_{k_4})
u_7 \cdot \epsilon (\lambda_{k_5}) u_8 \cdot \epsilon^* (\lambda_{k_6})
u_9 \cdot \epsilon^* (\lambda_{k_7}) 
{\rm tr} (\gammau_1 \gammau_2 \gammau_3 \gammau_4 \gammau_5 \gammau_6),
\end{displaymath}
\begin{displaymath}
u_7 \cdot u_8 \epsilon (\lambda_{k_1}) \cdot \epsilon (\lambda_{k_2})
\epsilon^* (\lambda_{k_3}) \cdot \epsilon^* (\lambda_{k_4})
u_9 \cdot \epsilon (\lambda_{k_5}) u_{10} \cdot \epsilon^* (\lambda_{k_6})
{\rm tr} (\gammau_1 \gammau_2 \gammau_3 \gammau_4 \gammau_5 \gammau_6),
\end{displaymath}
\begin{displaymath}
u_7 \cdot u_8 u_9 \cdot u_{10}
\epsilon (\lambda_{k_1}) \cdot \epsilon (\lambda_{k_2})
\epsilon^* (\lambda_{k_3}) \cdot \epsilon^* (\lambda_{k_4})
u_{11} \cdot \epsilon^* (\lambda_{k_5})
{\rm tr} (\gammau_1 \gammau_2 \gammau_3 \gammau_4 \gammau_5 \gammau_6),
\end{displaymath}
\begin{displaymath}
\epsilon (\lambda_{k_1}) \cdot \epsilon (\lambda_{k_2})
\epsilon (\lambda_{k_3}) \cdot \epsilon (\lambda_{k_4})
\epsilon^* (\lambda_{k_5}) \cdot \epsilon^* (\lambda_{k_6})
u_7 \cdot \epsilon^* (\lambda_{k_7}) u_8 \cdot \epsilon^* (\lambda_{k_8})
{\rm tr} (\gammau_1 \gammau_2 \gammau_3 \gammau_4 \gammau_5 \gammau_6),
\end{displaymath}
\begin{displaymath}
u_7 \cdot u_8 \epsilon (\lambda_{k_1}) \cdot \epsilon (\lambda_{k_2})
\epsilon^* (\lambda_{k_3}) \cdot \epsilon^* (\lambda_{k_4})
\epsilon^* (\lambda_{k_5}) \cdot \epsilon^* (\lambda_{k_6})
u_9 \cdot \epsilon (\lambda_{k_7})
{\rm tr} (\gammau_1 \gammau_2 \gammau_3 \gammau_4 \gammau_5 \gammau_6),
\end{displaymath}
\begin{displaymath}
u_5 \cdot u_6 u_7 \cdot u_8
\epsilon (\lambda_{k_1}) \cdot \epsilon (\lambda_{k_2})
\epsilon (\lambda_{k_3}) \cdot \epsilon (\lambda_{k_4})
\epsilon^* (\lambda_{k_5}) \cdot \epsilon^* (\lambda_{k_6})
\epsilon^* (\lambda_{k_7}) \cdot \epsilon^* (\lambda_{k_8})
{\rm tr} (\gammau_1 \gammau_2 \gammau_3 \gammau_4),
\end{displaymath}
\begin{displaymath}
u_5 \cdot \epsilon (\lambda_{k_1}) u_6 \cdot \epsilon (\lambda_{k_2})
u_7 \cdot \epsilon (\lambda_{k_3}) u_8 \cdot \epsilon^* (\lambda_{k_4})
u_9 \cdot \epsilon^* (\lambda_{k_5}) u_{10} \cdot \epsilon^* (\lambda_{k_6})
{\rm tr} (\gammau_1 \gammau_2 \gammau_3 \gammau_4),
\end{displaymath}
\begin{displaymath}
\epsilon^* (\lambda_{k_1}) \cdot \epsilon^* (\lambda_{k_2})
u_5 \cdot \epsilon (\lambda_{k_3}) u_6 \cdot \epsilon (\lambda_{k_4})
u_7 \cdot \epsilon (\lambda_{k_5}) u_8 \cdot \epsilon^* (\lambda_{k_6})
u_9 \cdot \epsilon^* (\lambda_{k_7}) 
{\rm tr} (\gammau_1 \gammau_2 \gammau_3 \gammau_4),
\end{displaymath}
\begin{displaymath}
u_5 \cdot u_6 \epsilon (\lambda_{k_1}) \cdot \epsilon (\lambda_{k_2})
u_7 \cdot \epsilon (\lambda_{k_3}) u_8 \cdot \epsilon^* (\lambda_{k_4})
u_9 \cdot \epsilon^* (\lambda_{k_5}) u_{10} \cdot \epsilon^* (\lambda_{k_6})
{\rm tr} (\gammau_1 \gammau_2 \gammau_3 \gammau_4),
\end{displaymath}
\begin{displaymath}
\epsilon (\lambda_{k_1}) \cdot \epsilon (\lambda_{k_2})
\epsilon^* (\lambda_{k_3}) \cdot \epsilon^* (\lambda_{k_4})
u_5 \cdot \epsilon (\lambda_{k_5}) u_6 \cdot \epsilon (\lambda_{k_6})
u_7 \cdot \epsilon^* (\lambda_{k_7}) u_8 \cdot \epsilon^* (\lambda_{k_8})
{\rm tr} (\gammau_1 \gammau_2 \gammau_3 \gammau_4),
\end{displaymath}
\begin{displaymath}
u_5 \cdot u_6 \epsilon (\lambda_{k_1}) \cdot \epsilon (\lambda_{k_2})
\epsilon^* (\lambda_{k_3}) \cdot \epsilon^* (\lambda_{k_4})
u_7 \cdot \epsilon (\lambda_{k_5}) u_8 \cdot \epsilon^* (\lambda_{k_6})
u_9 \cdot \epsilon^* (\lambda_{k_7})
{\rm tr} (\gammau_1 \gammau_2 \gammau_3 \gammau_4),
\end{displaymath}
\begin{displaymath}
u_5 \cdot u_6 u_7 \cdot u_8
\epsilon (\lambda_{k_1}) \cdot \epsilon (\lambda_{k_2})
\epsilon^* (\lambda_{k_3}) \cdot \epsilon^* (\lambda_{k_4})
u_9 \cdot \epsilon (\lambda_{k_5}) u_{10} \cdot \epsilon^* (\lambda_{k_6})
{\rm tr} (\gammau_1 \gammau_2 \gammau_3 \gammau_4),
\end{displaymath}
\begin{displaymath}
u_5 \cdot u_6 u_7 \cdot u_8
\epsilon (\lambda_{k_1}) \cdot \epsilon (\lambda_{k_2})
\epsilon^* (\lambda_{k_3}) \cdot \epsilon^* (\lambda_{k_4})
\epsilon^* (\lambda_{k_5}) \cdot \epsilon^* (\lambda_{k_6})
u_9 \cdot \epsilon (\lambda_{k_7}) 
{\rm tr} (\gammau_1 \gammau_2 \gammau_3 \gammau_4),
\end{displaymath}
\begin{displaymath}
u_5 \cdot u_6 \epsilon (\lambda_{k_1}) \cdot \epsilon (\lambda_{k_2})
\epsilon^* (\lambda_{k_3}) \cdot \epsilon^* (\lambda_{k_4})
\epsilon^* (\lambda_{k_5}) \cdot \epsilon^* (\lambda_{k_6})
u_7 \cdot \epsilon (\lambda_{k_7}) u_8 \cdot \epsilon (\lambda_{k_8})
{\rm tr} (\gammau_1 \gammau_2 \gammau_3 \gammau_4),
\end{displaymath}
where $u_h (h=1,\cdots, 16)$ are linear combinations of four-momenta of
external gluons, virtual gluons, external quarks, and virtual quarks, and
which do not include the $3 \times 3$ $SU(3)$ color matrices of the gluon-quark
vertex factor and the antisymmetric $SU(3)$ structure constants of the
triple-gluon vertex factor and the four-gluon vertex factor. The subscripts
$k_1, \cdots, k_8$ label the external gluons and take the values 1, 2, 4, and 
5. The traces of
products of Dirac matrices are calculated via the relation \cite{BD},
\begin{equation}
{\rm tr}(\gammac_1 \cdots \gammac_N)=
c_1 \cdot c_2 {\rm tr}(\gammac_3 \cdots \gammac_N)
-c_1 \cdot c_3 {\rm tr}(\gammac_2 \gammac_4 \cdots \gammac_N)
+ \cdots +c_1 \cdot c_N {\rm tr}(\gammac_2 \cdots \gammac_{N-1}),
\end{equation}
where $N$ is an even positive integer, and $c_h (h=1, \cdots , N)$ are
arbitrary four-vectors. The above types of structures turn into the following
twelve forms:
\begin{displaymath}
a_1 \cdot b_1 a_2 \cdot b_2 a_3 \cdot b_3 a_4 \cdot b_4 
\sum\limits_{\lambda_1 \lambda_2 \lambda_4 \lambda_5}
{\rm scalar~products~of~polarization~four-vectors},
\end{displaymath}
\begin{displaymath}
a_1 \cdot b_1 a_2 \cdot b_2 a_3 \cdot b_3 
\sum\limits_{\lambda_1 \lambda_2 \lambda_4 \lambda_5}
a_4 \cdot \epsilon (\lambda_i) \epsilon^*(\lambda_i) \cdot b_4  
\times {\rm scalar~products},
\end{displaymath}
\begin{displaymath}
a_1 \cdot b_1 a_2 \cdot b_2
\sum\limits_{\lambda_1 \lambda_2 \lambda_4 \lambda_5}
a_3 \cdot \epsilon (\lambda_i) \epsilon^*(\lambda_i) \cdot b_3
a_4 \cdot \epsilon (\lambda_j) \epsilon^*(\lambda_j) \cdot b_4
\times {\rm scalar~products},
\end{displaymath}
\begin{displaymath}
a_1 \cdot b_1 a_2 \cdot b_2 a_3 \cdot b_3 
\sum\limits_{\lambda_1 \lambda_2 \lambda_4 \lambda_5}
a_4 \cdot \epsilon (\lambda_i) 
\epsilon^* (\lambda_i) \cdot \epsilon (\lambda_j) 
\epsilon^*(\lambda_j) \cdot b_4
\times {\rm scalar~products},
\end{displaymath}
\begin{displaymath}
a_1 \cdot b_1
\sum\limits_{\lambda_1 \lambda_2 \lambda_4 \lambda_5}
a_2 \cdot \epsilon (\lambda_i) \epsilon^*(\lambda_i) \cdot b_2
a_3 \cdot \epsilon (\lambda_j) \epsilon^*(\lambda_j) \cdot b_3
a_4 \cdot \epsilon (\lambda_m) \epsilon^*(\lambda_m) \cdot b_4
\times {\rm scalar~products},
\end{displaymath}
\begin{displaymath}
a_1 \cdot b_1 a_2 \cdot b_2
\sum\limits_{\lambda_1 \lambda_2 \lambda_4 \lambda_5}
a_3 \cdot \epsilon (\lambda_i) \epsilon^*(\lambda_i) \cdot b_3
a_4 \cdot \epsilon (\lambda_j) 
\epsilon^* (\lambda_j) \cdot \epsilon (\lambda_m) 
\epsilon^*(\lambda_m) \cdot b_4
\times {\rm scalar~products},
\end{displaymath}
\begin{displaymath}
a_1 \cdot b_1 a_2 \cdot b_2 a_3 \cdot b_3 
\sum\limits_{\lambda_1 \lambda_2 \lambda_4 \lambda_5}
a_4 \cdot \epsilon (\lambda_i) 
\epsilon^* (\lambda_i) \cdot \epsilon (\lambda_j) 
\epsilon^* (\lambda_j) \cdot \epsilon (\lambda_m) 
\epsilon^*(\lambda_m) \cdot b_4
\times {\rm scalar~products},
\end{displaymath}
\begin{displaymath}
\sum\limits_{\lambda_1 \lambda_2 \lambda_4 \lambda_5}
a_1 \cdot \epsilon (\lambda_i) \epsilon^*(\lambda_i) \cdot b_1
a_2 \cdot \epsilon (\lambda_j) \epsilon^*(\lambda_j) \cdot b_2
a_3 \cdot \epsilon (\lambda_m) \epsilon^*(\lambda_m) \cdot b_3
a_4 \cdot \epsilon (\lambda_n) \epsilon^*(\lambda_n) \cdot b_4,
\end{displaymath}
\begin{displaymath}
a_1 \cdot b_1
\sum\limits_{\lambda_1 \lambda_2 \lambda_4 \lambda_5}
a_2 \cdot \epsilon (\lambda_i) \epsilon^*(\lambda_i) \cdot b_2
a_3 \cdot \epsilon (\lambda_j) \epsilon^*(\lambda_j) \cdot b_3
a_4 \cdot \epsilon (\lambda_m) 
\epsilon^* (\lambda_m) \cdot \epsilon (\lambda_n) 
\epsilon^*(\lambda_n) \cdot b_4,
\end{displaymath}
\begin{displaymath}
a_1 \cdot b_1 a_2 \cdot b_2
\sum\limits_{\lambda_1 \lambda_2 \lambda_4 \lambda_5}
a_3 \cdot \epsilon (\lambda_i) 
\epsilon^* (\lambda_i) \cdot \epsilon (\lambda_j) 
\epsilon^*(\lambda_j) \cdot b_3
a_4 \cdot \epsilon (\lambda_m) 
\epsilon^* (\lambda_m) \cdot \epsilon (\lambda_n) 
\epsilon^*(\lambda_n) \cdot b_4,
\end{displaymath}
\begin{displaymath}
a_1 \cdot b_1 a_2 \cdot b_2
\sum\limits_{\lambda_1 \lambda_2 \lambda_4 \lambda_5}
a_3 \cdot \epsilon (\lambda_i) \epsilon^*(\lambda_i) \cdot b_3
a_4 \cdot \epsilon (\lambda_j) 
\epsilon^* (\lambda_j) \cdot \epsilon (\lambda_m) 
\epsilon^* (\lambda_m) \cdot \epsilon (\lambda_n) 
\epsilon^*(\lambda_n) \cdot b_4,
\end{displaymath}
\begin{displaymath}
a_1 \cdot b_1 a_2 \cdot b_2 a_3 \cdot b_3
\sum\limits_{\lambda_1 \lambda_2 \lambda_4 \lambda_5}
a_4 \cdot \epsilon (\lambda_i) 
\epsilon^* (\lambda_i) \cdot \epsilon (\lambda_j) 
\epsilon^* (\lambda_j) \cdot \epsilon (\lambda_m) 
\epsilon^* (\lambda_m) \cdot \epsilon (\lambda_n) 
\epsilon^*(\lambda_n) \cdot b_4,
\end{displaymath}
where $a_h$ and $b_h$ $(h=1,2,3,4)$ are linear combinations of four-momenta;
the subscripts $i$, $j$, $m$, and $n$ label the external gluons, take the
values 1, 2, 4, and 5, and do not equal each other.
For an external gluon in the physical transverse state with four-momentum $p_i$
and helicity $\lambda_i$, we use the projection \cite{field},
\begin{equation}
\sum\limits_{\lambda_i} \epsilon_\mu (\lambda_i) \epsilon^*_\nu (\lambda_i)
= -g_{\mu \nu} + \frac {w_\mu p_{i\nu} + w_\nu p_{i\mu}}{w \cdot p_i}
-\frac {w^2p_{i\mu} p_{i\nu}}{(w \cdot p_i)^2},
\end{equation}
where $w$ is an arbitrary four-vector. A convenient choice for $w$ is $w=p_3$
with $p^2_3=0$. Eq. (2) leads to
\begin{equation}
p_i^{\mu}\sum\limits_{\lambda_i}
\epsilon_\mu (\lambda_i) \epsilon^*_\nu (\lambda_i) =0,
\end{equation}
\begin{equation}
p_i^{\nu}\sum\limits_{\lambda_i}
\epsilon_\mu (\lambda_i) \epsilon^*_\nu (\lambda_i) =0.
\end{equation}
The scalar products of polarization four-vectors in the first seven of the
above twelve forms take the following expressions with the values -2 and 2:
\begin{equation}
\sum\limits_{\lambda_i} \epsilon (\lambda_i) \cdot \epsilon^* (\lambda_i) = -2,
\end{equation}
\begin{equation}
\sum\limits_{\lambda_i \lambda_j}
\epsilon (\lambda_i) \cdot \epsilon (\lambda_j)
\epsilon^* (\lambda_i) \cdot \epsilon^* (\lambda_j)
=\sum\limits_{\lambda_i \lambda_j}
\epsilon (\lambda_i) \cdot \epsilon^* (\lambda_j)
\epsilon^* (\lambda_i) \cdot \epsilon (\lambda_j)= 2,
\end{equation}
\begin{equation}
\sum\limits_{\lambda_i \lambda_j \lambda_k}
\epsilon^* (\lambda_i) \cdot \epsilon^* (\lambda_j)
\epsilon (\lambda_j) \cdot \epsilon^* (\lambda_k)
\epsilon (\lambda_k) \cdot \epsilon (\lambda_i)=-2,
\end{equation}
\begin{equation}
\sum\limits_{\lambda_i \lambda_j \lambda_k}
\epsilon^* (\lambda_i) \cdot \epsilon^* (\lambda_j)
\epsilon (\lambda_j) \cdot \epsilon (\lambda_k)
\epsilon^* (\lambda_k) \cdot \epsilon (\lambda_i)=-2,
\end{equation}
\begin{equation}
\sum\limits_{\lambda_i \lambda_j \lambda_k}
\epsilon^* (\lambda_i) \cdot \epsilon (\lambda_j)
\epsilon^* (\lambda_j) \cdot \epsilon^* (\lambda_k)
\epsilon (\lambda_k) \cdot \epsilon (\lambda_i)=-2,
\end{equation}
\begin{equation}
\sum\limits_{\lambda_i \lambda_j \lambda_k}
\epsilon^* (\lambda_i) \cdot \epsilon (\lambda_j)
\epsilon^* (\lambda_j) \cdot \epsilon (\lambda_k)
\epsilon^* (\lambda_k) \cdot \epsilon (\lambda_i)=-2,
\end{equation}
\begin{equation}
\sum\limits_{\lambda_i \lambda_j \lambda_k}
\epsilon (\lambda_i) \cdot \epsilon^* (\lambda_j)
\epsilon (\lambda_j) \cdot \epsilon^* (\lambda_k)
\epsilon (\lambda_k) \cdot \epsilon^* (\lambda_i)=-2,
\end{equation}
\begin{equation}
\sum\limits_{\lambda_i \lambda_j \lambda_k}
\epsilon (\lambda_i) \cdot \epsilon^* (\lambda_j)
\epsilon (\lambda_j) \cdot \epsilon (\lambda_k)
\epsilon^* (\lambda_k) \cdot \epsilon^* (\lambda_i)=-2,
\end{equation}
\begin{equation}
\sum\limits_{\lambda_i \lambda_j \lambda_k}
\epsilon (\lambda_i) \cdot \epsilon (\lambda_j)
\epsilon^* (\lambda_j) \cdot \epsilon^* (\lambda_k)
\epsilon (\lambda_k) \cdot \epsilon^* (\lambda_i)= -2,
\end{equation}
\begin{equation}
\sum\limits_{\lambda_i \lambda_j \lambda_k}
\epsilon (\lambda_i) \cdot \epsilon (\lambda_j)
\epsilon^* (\lambda_j) \cdot \epsilon (\lambda_k)
\epsilon^* (\lambda_k) \cdot \epsilon^* (\lambda_i)= -2,
\end{equation}
\begin{equation}
\sum\limits_{\lambda_i \lambda_j \lambda_m \lambda_n}
\epsilon^* (\lambda_i) \cdot \epsilon^* (\lambda_j)
\epsilon (\lambda_j) \cdot \epsilon^* (\lambda_m)
\epsilon (\lambda_m) \cdot \epsilon^* (\lambda_n)
\epsilon (\lambda_n) \cdot \epsilon (\lambda_i)=2,
\end{equation}
\begin{equation}
\sum\limits_{\lambda_i \lambda_j \lambda_m \lambda_n}
\epsilon^* (\lambda_i) \cdot \epsilon^* (\lambda_j)
\epsilon (\lambda_j) \cdot \epsilon^* (\lambda_m)
\epsilon (\lambda_m) \cdot \epsilon (\lambda_n)
\epsilon^* (\lambda_n) \cdot \epsilon (\lambda_i)=2,
\end{equation}
\begin{equation}
\sum\limits_{\lambda_i \lambda_j \lambda_m \lambda_n}
\epsilon^* (\lambda_i) \cdot \epsilon^* (\lambda_j)
\epsilon (\lambda_j) \cdot \epsilon (\lambda_m)
\epsilon^* (\lambda_m) \cdot \epsilon^* (\lambda_n)
\epsilon (\lambda_n) \cdot \epsilon (\lambda_i)=2,
\end{equation}
\begin{equation}
\sum\limits_{\lambda_i \lambda_j \lambda_m \lambda_n}
\epsilon^* (\lambda_i) \cdot \epsilon^* (\lambda_j)
\epsilon (\lambda_j) \cdot \epsilon (\lambda_m)
\epsilon^* (\lambda_m) \cdot \epsilon (\lambda_n)
\epsilon^* (\lambda_n) \cdot \epsilon (\lambda_i)=2,
\end{equation}
\begin{equation}
\sum\limits_{\lambda_i \lambda_j \lambda_m \lambda_n}
\epsilon^* (\lambda_i) \cdot \epsilon (\lambda_j)
\epsilon^* (\lambda_j) \cdot \epsilon^* (\lambda_m)
\epsilon (\lambda_m) \cdot \epsilon^* (\lambda_n)
\epsilon (\lambda_n) \cdot \epsilon (\lambda_i)=2,
\end{equation}
\begin{equation}
\sum\limits_{\lambda_i \lambda_j \lambda_m \lambda_n}
\epsilon^* (\lambda_i) \cdot \epsilon (\lambda_j)
\epsilon^* (\lambda_j) \cdot \epsilon^* (\lambda_m)
\epsilon (\lambda_m) \cdot \epsilon (\lambda_n)
\epsilon^* (\lambda_n) \cdot \epsilon (\lambda_i)=2,
\end{equation}
\begin{equation}
\sum\limits_{\lambda_i \lambda_j \lambda_m \lambda_n}
\epsilon^* (\lambda_i) \cdot \epsilon (\lambda_j)
\epsilon^* (\lambda_j) \cdot \epsilon (\lambda_m)
\epsilon^* (\lambda_m) \cdot \epsilon^* (\lambda_n)
\epsilon (\lambda_n) \cdot \epsilon (\lambda_i)=2,
\end{equation}
\begin{equation}
\sum\limits_{\lambda_i \lambda_j \lambda_m \lambda_n}
\epsilon^* (\lambda_i) \cdot \epsilon (\lambda_j)
\epsilon^* (\lambda_j) \cdot \epsilon (\lambda_m)
\epsilon^* (\lambda_m) \cdot \epsilon (\lambda_n)
\epsilon^* (\lambda_n) \cdot \epsilon (\lambda_i)=2,
\end{equation}
\begin{equation}
\sum\limits_{\lambda_i \lambda_j \lambda_m \lambda_n}
\epsilon (\lambda_i) \cdot \epsilon^* (\lambda_j)
\epsilon (\lambda_j) \cdot \epsilon^* (\lambda_m)
\epsilon (\lambda_m) \cdot \epsilon^* (\lambda_n)
\epsilon (\lambda_n) \cdot \epsilon^* (\lambda_i)=2,
\end{equation}
\begin{equation}
\sum\limits_{\lambda_i \lambda_j \lambda_m \lambda_n}
\epsilon (\lambda_i) \cdot \epsilon^* (\lambda_j)
\epsilon (\lambda_j) \cdot \epsilon^* (\lambda_m)
\epsilon (\lambda_m) \cdot \epsilon (\lambda_n)
\epsilon^* (\lambda_n) \cdot \epsilon^* (\lambda_i)=2,
\end{equation}
\begin{equation}
\sum\limits_{\lambda_i \lambda_j \lambda_m \lambda_n}
\epsilon (\lambda_i) \cdot \epsilon^* (\lambda_j)
\epsilon (\lambda_j) \cdot \epsilon (\lambda_m)
\epsilon^* (\lambda_m) \cdot \epsilon^* (\lambda_n)
\epsilon (\lambda_n) \cdot \epsilon^* (\lambda_i)=2,
\end{equation}
\begin{equation}
\sum\limits_{\lambda_i \lambda_j \lambda_m \lambda_n}
\epsilon (\lambda_i) \cdot \epsilon^* (\lambda_j)
\epsilon (\lambda_j) \cdot \epsilon (\lambda_m)
\epsilon^* (\lambda_m) \cdot \epsilon (\lambda_n)
\epsilon^* (\lambda_n) \cdot \epsilon^* (\lambda_i)=2,
\end{equation}
\begin{equation}
\sum\limits_{\lambda_i \lambda_j \lambda_m \lambda_n}
\epsilon (\lambda_i) \cdot \epsilon (\lambda_j)
\epsilon^* (\lambda_j) \cdot \epsilon^* (\lambda_m)
\epsilon (\lambda_m) \cdot \epsilon^* (\lambda_n)
\epsilon (\lambda_n) \cdot \epsilon^* (\lambda_i)=2,
\end{equation}
\begin{equation}
\sum\limits_{\lambda_i \lambda_j \lambda_m \lambda_n}
\epsilon (\lambda_i) \cdot \epsilon (\lambda_j)
\epsilon^* (\lambda_j) \cdot \epsilon^* (\lambda_m)
\epsilon (\lambda_m) \cdot \epsilon (\lambda_n)
\epsilon^* (\lambda_n) \cdot \epsilon^* (\lambda_i)=2,
\end{equation}
\begin{equation}
\sum\limits_{\lambda_i \lambda_j \lambda_m \lambda_n}
\epsilon (\lambda_i) \cdot \epsilon (\lambda_j)
\epsilon^* (\lambda_j) \cdot \epsilon (\lambda_m)
\epsilon^* (\lambda_m) \cdot \epsilon^* (\lambda_n)
\epsilon (\lambda_n) \cdot \epsilon^* (\lambda_i)=2,
\end{equation}
\begin{equation}
\sum\limits_{\lambda_i \lambda_j \lambda_m \lambda_n}
\epsilon (\lambda_i) \cdot \epsilon (\lambda_j)
\epsilon^* (\lambda_j) \cdot \epsilon (\lambda_m)
\epsilon^* (\lambda_m) \cdot \epsilon (\lambda_n)
\epsilon^* (\lambda_n) \cdot \epsilon^* (\lambda_i)=2,
\end{equation}
where the subscripts $i$, $j$, $k$, $m$, and $n$ label the external gluons, 
take the values 1, 2, 4, and 5, and do
not equal each other. Aside from the scalar products of polarization
four-vectors, the other expressions encountered in the twelve forms before Eq.
(2) are expressed in terms of scalar products of four-momenta,
\begin{equation}
\sum\limits_{\lambda_i}
a \cdot \epsilon (\lambda_i) \epsilon^* (\lambda_i) \cdot b 
= -a \cdot b + \frac{a \cdot p_3 b \cdot p_i + b \cdot p_3 a \cdot p_i}
{p_3 \cdot p_i},
\end{equation}
\begin{eqnarray}
& & \sum\limits_{\lambda_i \lambda_j}
a \cdot \epsilon (\lambda_i) \epsilon^*(\lambda_i) \cdot \epsilon (\lambda_j)
\epsilon^*(\lambda_j) \cdot b
= \sum\limits_{\lambda_i \lambda_j}
a \cdot \epsilon^*(\lambda_i) \epsilon (\lambda_i) \cdot \epsilon^*(\lambda_j)
\epsilon (\lambda_j) \cdot b
                                     \nonumber    \\
& = & \sum\limits_{\lambda_i \lambda_j}
a \cdot \epsilon^*(\lambda_i) \epsilon (\lambda_i) \cdot \epsilon (\lambda_j)
\epsilon^*(\lambda_j) \cdot b
= \sum\limits_{\lambda_i \lambda_j}
a \cdot \epsilon (\lambda_i) \epsilon^*(\lambda_i) \cdot \epsilon^*(\lambda_j)
\epsilon (\lambda_j) \cdot b
                                     \nonumber    \\
& = & a \cdot b - \frac {b \cdot p_3 a \cdot p_j}{p_3 \cdot p_j}
- \frac {a \cdot p_3 b \cdot p_i}{p_3 \cdot p_i}
+ \frac {a \cdot p_3 p_i \cdot p_j b \cdot p_3}{p_3 \cdot p_i p_3 \cdot p_j},
                                     \nonumber    \\
\end{eqnarray}
\begin{eqnarray}
& & \sum\limits_{\lambda_i \lambda_j \lambda_k}
a \cdot \epsilon^*(\lambda_i) \epsilon (\lambda_i) \cdot \epsilon^*(\lambda_j)
\epsilon (\lambda_j) \cdot \epsilon^*(\lambda_k) \epsilon (\lambda_k) \cdot b
                                     \nonumber    \\
& = & \sum\limits_{\lambda_i \lambda_j \lambda_k}
a \cdot \epsilon^*(\lambda_i) \epsilon (\lambda_i) \cdot \epsilon^*(\lambda_j)
\epsilon (\lambda_j) \cdot \epsilon (\lambda_k) \epsilon^*(\lambda_k) \cdot b
                                     \nonumber    \\
& = & \sum\limits_{\lambda_i \lambda_j \lambda_k}
a \cdot \epsilon^*(\lambda_i) \epsilon (\lambda_i) \cdot \epsilon (\lambda_j)
\epsilon^*(\lambda_j) \cdot \epsilon^*(\lambda_k) \epsilon (\lambda_k) \cdot b
                                     \nonumber    \\
& = & \sum\limits_{\lambda_i \lambda_j \lambda_k}
a \cdot \epsilon^*(\lambda_i) \epsilon (\lambda_i) \cdot \epsilon (\lambda_j)
\epsilon^*(\lambda_j) \cdot \epsilon (\lambda_k) \epsilon^*(\lambda_k) \cdot b
                                     \nonumber    \\
& = & \sum\limits_{\lambda_i \lambda_j \lambda_k}
a \cdot \epsilon (\lambda_i) \epsilon^*(\lambda_i) \cdot \epsilon^*(\lambda_j)
\epsilon (\lambda_j) \cdot \epsilon^*(\lambda_k) \epsilon (\lambda_k) \cdot b
                                     \nonumber    \\
& = & \sum\limits_{\lambda_i \lambda_j \lambda_k}
a \cdot \epsilon (\lambda_i) \epsilon^*(\lambda_i) \cdot \epsilon^*(\lambda_j)
\epsilon (\lambda_j) \cdot \epsilon (\lambda_k) \epsilon^*(\lambda_k) \cdot b
                                     \nonumber    \\
& = & \sum\limits_{\lambda_i \lambda_j \lambda_k}
a \cdot \epsilon (\lambda_i) \epsilon^*(\lambda_i) \cdot \epsilon(\lambda_j)
\epsilon^*(\lambda_j) \cdot \epsilon^*(\lambda_k) \epsilon (\lambda_k) \cdot b
                                     \nonumber    \\
& = & \sum\limits_{\lambda_i \lambda_j \lambda_k}
a \cdot \epsilon (\lambda_i) \epsilon^*(\lambda_i) \cdot \epsilon (\lambda_j)
\epsilon^*(\lambda_j) \cdot \epsilon (\lambda_k) \epsilon^*(\lambda_k) \cdot b
                                     \nonumber    \\
& = & - a \cdot b + \frac {b \cdot p_3 a \cdot p_k}{p_3 \cdot p_k}
+ \frac {a \cdot p_3 b \cdot p_i}{p_3 \cdot p_i}
- \frac {a \cdot p_3 p_i \cdot p_k b \cdot p_3}{p_3 \cdot p_i p_3 \cdot p_k},
                                     \nonumber    \\
\end{eqnarray}
which is independent of $p_j$, and
\begin{eqnarray}
& & \sum\limits_{\lambda_i \lambda_j \lambda_m \lambda_n}
a \cdot \epsilon^* (\lambda_i) \epsilon (\lambda_i) \cdot \epsilon^*(\lambda_j)
\epsilon (\lambda_j) \cdot \epsilon^*(\lambda_m) 
\epsilon (\lambda_m) \cdot \epsilon^* (\lambda_n) 
\epsilon (\lambda_n) \cdot b
                                     \nonumber    \\
& = & \sum\limits_{\lambda_i \lambda_j \lambda_m \lambda_n}
a \cdot \epsilon^*(\lambda_i) \epsilon (\lambda_i) \cdot \epsilon^*(\lambda_j)
\epsilon (\lambda_j) \cdot \epsilon^* (\lambda_m) 
\epsilon (\lambda_m) \cdot \epsilon (\lambda_n) 
\epsilon^*(\lambda_n) \cdot b
                                     \nonumber    \\
& = & \sum\limits_{\lambda_i \lambda_j \lambda_m \lambda_n}
a \cdot \epsilon^*(\lambda_i) \epsilon (\lambda_i) \cdot \epsilon^*(\lambda_j)
\epsilon (\lambda_j) \cdot \epsilon (\lambda_m) 
\epsilon^* (\lambda_m) \cdot \epsilon^* (\lambda_n) 
\epsilon (\lambda_n) \cdot b
                                     \nonumber    \\
& = & \sum\limits_{\lambda_i \lambda_j \lambda_m \lambda_n}
a \cdot \epsilon^*(\lambda_i) \epsilon (\lambda_i) \cdot \epsilon^*(\lambda_j)
\epsilon (\lambda_j) \cdot \epsilon (\lambda_m) 
\epsilon^*(\lambda_m) \cdot \epsilon (\lambda_n) 
\epsilon^*(\lambda_n) \cdot b
                                     \nonumber    \\
& = & \sum\limits_{\lambda_i \lambda_j \lambda_m \lambda_n}
a \cdot \epsilon^*(\lambda_i) \epsilon (\lambda_i) \cdot \epsilon (\lambda_j)
\epsilon^*(\lambda_j) \cdot \epsilon^*(\lambda_m) 
\epsilon (\lambda_m) \cdot \epsilon^*(\lambda_n) 
\epsilon (\lambda_n) \cdot b
                                     \nonumber    \\
& = & \sum\limits_{\lambda_i \lambda_j \lambda_m \lambda_n}
a \cdot \epsilon^* (\lambda_i) \epsilon (\lambda_i) \cdot \epsilon (\lambda_j)
\epsilon^*(\lambda_j) \cdot \epsilon^* (\lambda_m) 
\epsilon (\lambda_m) \cdot \epsilon (\lambda_n) 
\epsilon^*(\lambda_n) \cdot b
                                     \nonumber    \\
& = & \sum\limits_{\lambda_i \lambda_j \lambda_m \lambda_n}
a \cdot \epsilon^*(\lambda_i) \epsilon (\lambda_i) \cdot \epsilon (\lambda_j)
\epsilon^*(\lambda_j) \cdot \epsilon (\lambda_m) 
\epsilon^*(\lambda_m) \cdot \epsilon^* (\lambda_n) 
\epsilon (\lambda_n) \cdot b
                                     \nonumber    \\
& = & \sum\limits_{\lambda_i \lambda_j \lambda_m \lambda_n}
a \cdot \epsilon^*(\lambda_i) \epsilon (\lambda_i) \cdot \epsilon (\lambda_j)
\epsilon^*(\lambda_j) \cdot \epsilon (\lambda_m) 
\epsilon^*(\lambda_m) \cdot \epsilon (\lambda_n) 
\epsilon^*(\lambda_n) \cdot b
                                     \nonumber    \\
& = & \sum\limits_{\lambda_i \lambda_j \lambda_m \lambda_n}
a \cdot \epsilon (\lambda_i) \epsilon^*(\lambda_i) \cdot \epsilon^* (\lambda_j)
\epsilon (\lambda_j) \cdot \epsilon^* (\lambda_m) 
\epsilon (\lambda_m) \cdot \epsilon^* (\lambda_n) 
\epsilon (\lambda_n) \cdot b
                                     \nonumber    \\
& = & \sum\limits_{\lambda_i \lambda_j \lambda_m \lambda_n}
a \cdot \epsilon (\lambda_i) \epsilon^*(\lambda_i) \cdot \epsilon^*(\lambda_j)
\epsilon (\lambda_j) \cdot \epsilon^* (\lambda_m) 
\epsilon (\lambda_m) \cdot \epsilon (\lambda_n) 
\epsilon^*(\lambda_n) \cdot b
                                     \nonumber    \\
& = & \sum\limits_{\lambda_i \lambda_j \lambda_m \lambda_n}
a \cdot \epsilon (\lambda_i) \epsilon^*(\lambda_i) \cdot \epsilon^*(\lambda_j)
\epsilon (\lambda_j) \cdot \epsilon (\lambda_m) 
\epsilon^*(\lambda_m) \cdot \epsilon^*(\lambda_n) 
\epsilon (\lambda_n) \cdot b
                                     \nonumber    \\
& = & \sum\limits_{\lambda_i \lambda_j \lambda_m \lambda_n}
a \cdot \epsilon (\lambda_i) \epsilon^*(\lambda_i) \cdot \epsilon^*(\lambda_j)
\epsilon (\lambda_j) \cdot \epsilon (\lambda_m) 
\epsilon^*(\lambda_m) \cdot \epsilon (\lambda_n) 
\epsilon^*(\lambda_n) \cdot b
                                     \nonumber    \\
& = & \sum\limits_{\lambda_i \lambda_j \lambda_m \lambda_n}
a \cdot \epsilon (\lambda_i) \epsilon^*(\lambda_i) \cdot \epsilon (\lambda_j)
\epsilon^*(\lambda_j) \cdot \epsilon^*(\lambda_m) 
\epsilon (\lambda_m) \cdot \epsilon^* (\lambda_n) 
\epsilon (\lambda_n) \cdot b
                                     \nonumber    \\
& = & \sum\limits_{\lambda_i \lambda_j \lambda_m \lambda_n}
a \cdot \epsilon (\lambda_i) \epsilon^*(\lambda_i) \cdot \epsilon (\lambda_j)
\epsilon^*(\lambda_j) \cdot \epsilon^* (\lambda_m) 
\epsilon (\lambda_m) \cdot \epsilon (\lambda_n) 
\epsilon^*(\lambda_n) \cdot b
                                     \nonumber    \\
& = & \sum\limits_{\lambda_i \lambda_j \lambda_m \lambda_n}
a \cdot \epsilon (\lambda_i) \epsilon^*(\lambda_i) \cdot \epsilon (\lambda_j)
\epsilon^*(\lambda_j) \cdot \epsilon (\lambda_m) 
\epsilon^*(\lambda_m) \cdot \epsilon^*(\lambda_n) 
\epsilon (\lambda_n) \cdot b
                                     \nonumber    \\
& = & \sum\limits_{\lambda_i \lambda_j \lambda_m \lambda_n}
a \cdot \epsilon (\lambda_i) \epsilon^*(\lambda_i) \cdot \epsilon (\lambda_j)
\epsilon^*(\lambda_j) \cdot \epsilon (\lambda_m) 
\epsilon^*(\lambda_m) \cdot \epsilon (\lambda_n) 
\epsilon^*(\lambda_n) \cdot b
                                     \nonumber    \\
& = & a \cdot b - \frac {b \cdot p_3 a \cdot p_n}{p_3 \cdot p_n}
- \frac {a \cdot p_3 b \cdot p_i}{p_3 \cdot p_i}
+ \frac {a \cdot p_3 p_i \cdot p_n b \cdot p_3}{p_3 \cdot p_i p_3 \cdot p_n},
                                     \nonumber    \\
\end{eqnarray}
which is independent of $p_j$ and $p_m$. In Eqs. (31)-(34)
$a$ and $b$ are linear combinations of four-momenta.

Every gluon-quark vertex factor has a $SU(3)$ color matrix, every triple-gluon
vertex factor has a $SU(3)$ structure constant, and every four-gluon vertex
factor has two $SU(3)$ structure constants. From a diagram in the first class
to a diagram in the fourth class, the number of gluon-quark vertices decreases
and the number of triple-gluon vertices increases. 
While the number of gluon-quark
vertices decreases by one, the number of triple-gluon vertices increases by 
one. Therefore, in a diagram's amplitude the total number of the $SU(3)$ color
matrices and the $SU(3)$ structure constants is four. An individually squared
amplitude or an interference term involves the product of $n$ ($0 \leq n \leq 
6$)  $SU(3)$ structure constants and the trace of the product of $8-n$ $SU(3)$
color matrices. Fortran code is made to calculate the products.

Including the average over the spin and color states of the two initial gluons
and the initial quark, finally, the spin- and color-summed squared amplitude
for a diagram and the spin- and color-summed interference term of two
diagrams are given as
functions of the scalar products of the four-momenta of external gluons and
external quarks. By means of
$s_{12}=(p_1+p_2)^2$, $s_{23}=(p_2+p_3)^2$, $s_{31}=(p_3+p_1)^2$,
$u_{15}=(p_1-p_5)^2$, $u_{16}=(p_1-p_6)^2$, $u_{24}=(p_2-p_4)^2$,
$u_{26}=(p_2-p_6)^2$, $u_{34}=(p_3-p_4)^2$, and $u_{35}=(p_3-p_5)^2$, the 123
individually squared amplitudes and the interference terms of the 123 Feynman
diagrams are expressed in terms of the nine independent variables $s_{12}$,
$s_{23}$, $s_{31}$, $u_{15}$, $u_{16}$, $u_{24}$, $u_{26}$, $u_{34}$, and
$u_{35}$. We denote by 
$\mid {\cal M}_{ggq \to ggq} \mid^2$ the squared amplitude for the elastic
$ggq$ scattering, which is the sum of the individually squared
amplitudes and the interference terms. The
amplitude ${\cal M}_{gg\bar {q} \to gg\bar {q}}$ for the elastic $gg\bar q$
scattering satisfies the relation
$\mid {\cal M}_{gg\bar {q} \to gg\bar {q}} \mid^2 = 
\mid {\cal M}_{ggq \to ggq} \mid^2$.

\vspace{0.5cm}
\leftline{\bf IV. TRANSPORT EQUATIONS}
\vspace{0.5cm}

In the last section we have obtained the squared amplitudes for the elastic
$ggq$ scattering and the elastic $gg\bar q$ scattering.
The other types of elastic 3-to-3 scattering have already been studied in Refs.
\cite{XSCZ,XMCW,XX,XPW} for the evolution of quark-gluon matter. We assume that
quark-gluon matter consists of gluons and equal amounts of up quarks, down
quarks, up antiquarks, and down antiquarks. Elastic 2-to-2 scattering and
elastic 3-to-3 scattering involve quarks and antiquarks on an equal footing,
and antiquark matter takes the same evolution as quark matter. Let
${\cal M}_{gqq' \to gqq'}$ be the amplitude for elastic gluon-quark-quark
scattering, and let ${\cal M}_{{\bar q}{\bar q}' \to {\bar q}{\bar q}'}$ be the
amplitude for elastic antiquark-antiquark scattering, where $q$ $(q')$ is the 
up or down quark, and $\bar {q}$ $(\bar {q}')$ is the up or down antiquark. 
In the same way we understand the notations ${\cal M}_{ggg \to ggg}$,
${\cal M}_{g{\bar q}{\bar q}' \to g{\bar q}{\bar q}'}$,
${\cal M}_{gq{\bar q}' \to gq{\bar q}'}$, ${\cal M}_{qqq' \to qqq'}$, 
${\cal M}_{{\bar q}{\bar q}{\bar q}' \to {\bar q}{\bar q}{\bar q}'}$,
${\cal M}_{qq{\bar q}' \to qq{\bar q}'}$,
${\cal M}_{qq'{\bar q} \to qq'{\bar q}}$,
${\cal M}_{q'{\bar q}{\bar q} \to q'{\bar q}{\bar q}}$,
${\cal M}_{q{\bar q}{\bar q}' \to q{\bar q}{\bar q}'}$,
${\cal M}_{gg \to gg}$, ${\cal M}_{gq \to gq}$,
${\cal M}_{g{\bar q} \to g{\bar q}}$,
${\cal M}_{qq' \to qq'}$, and 
${\cal M}_{q{\bar q}' \to q{\bar q}'}$. While quark-gluon matter is governed by
all types of elastic 2-to-2 scattering and elastic 3-to-3 scattering,
the transport equation for gluon matter is
\begin{eqnarray}
\frac {\partial f_{g1}}{\partial t}
& + & \vec {\rm v}_1 \cdot \vec {\nabla}_{\vec {r}} f_{g1}
= -\frac {1}{2E_1} \int \frac {d^3p_2}{(2\pi)^32E_2}
\frac {d^3p_3}{(2\pi)^32E_3} \frac {d^3p_4}{(2\pi)^32E_4}
(2\pi)^4 \delta^4(p_1+p_2-p_3-p_4)
         \nonumber    \\
& & \times \left\{ \frac {g_G}{2} \mid {\cal M}_{gg \to gg} \mid^2
[f_{g1}f_{g2}(1+f_{g3})(1+f_{g4})-f_{g3}f_{g4}(1+f_{g1})(1+f_{g2})]  \right.
         \nonumber    \\
& & + g_Q ( \mid {\cal M}_{gu \to gu} \mid^2
+ \mid {\cal M}_{gd \to gd} \mid^2
+ \mid {\cal M}_{g\bar {u} \to g\bar {u}} \mid^2
+ \mid {\cal M}_{g\bar {d} \to g\bar {d}} \mid^2 )
         \nonumber    \\
& & \left.
\times [f_{g1}f_{q2}(1+f_{g3})(1-f_{q4})-f_{g3}f_{q4}(1+f_{g1})(1-f_{q2})]
    \right\}
         \nonumber    \\
& & -\frac {1}{2E_1} \int \frac {d^3p_2}{(2\pi)^32E_2}
\frac {d^3p_3}{(2\pi)^32E_3} \frac {d^3p_4}{(2\pi)^32E_4}
\frac {d^3p_5}{(2\pi)^32E_5} \frac {d^3p_6}{(2\pi)^32E_6}
         \nonumber    \\
& & \times (2\pi)^4 \delta^4(p_1+p_2+p_3-p_4-p_5-p_6) 
\left\{ \frac {g_G^2}{12} \mid {\cal M}_{ggg \to ggg} \mid^2  \right.
         \nonumber    \\
& & \times [f_{g1}f_{g2}f_{g3}(1+f_{g4})(1+f_{g5})(1+f_{g6})-
f_{g4}f_{g5}f_{g6}(1+f_{g1})(1+f_{g2})(1+f_{g3})]
         \nonumber    \\
& & + \frac {g_Gg_Q}{2} ( \mid {\cal M}_{ggu \to ggu} \mid^2
+\mid {\cal M}_{ggd \to ggd} \mid^2
+\mid {\cal M}_{gg\bar {u} \to gg\bar {u}} \mid^2 
+\mid {\cal M}_{gg\bar {d} \to gg\bar {d}} \mid^2 )
         \nonumber    \\
& & \times [f_{g1}f_{g2}f_{q3}(1+f_{g4})(1+f_{g5})(1-f_{q6})
-f_{g4}f_{g5}f_{q6}(1+f_{g1})(1+f_{g2})(1-f_{q3})]
         \nonumber    \\
& & + g_Q^2 [\frac {1}{4} \mid {\cal M}_{guu \to guu} \mid^2
+\frac {1}{2} ( \mid {\cal M}_{gud \to gud} \mid^2
              + \mid {\cal M}_{gdu \to gdu} \mid^2 )
+\frac {1}{4} \mid {\cal M}_{gdd \to gdd} \mid^2
         \nonumber    \\
& & + \mid {\cal M}_{gu\bar {u} \to gu\bar {u}} \mid^2
    + \mid {\cal M}_{gu\bar {d} \to gu\bar {d}} \mid^2
    + \mid {\cal M}_{gd\bar {u} \to gd\bar {u}} \mid^2
    + \mid {\cal M}_{gd\bar {d} \to gd\bar {d}} \mid^2
         \nonumber        \\
& & +\frac {1}{4} \mid {\cal M}_{g\bar {u}\bar {u}
                             \to g\bar {u}\bar {u}} \mid^2
    +\frac {1}{2} ( \mid {\cal M}_{g\bar {u}\bar {d}
                               \to g\bar {u}\bar {d}} \mid^2 
    + \mid {\cal M}_{g\bar {d}\bar {u} \to g\bar {d}\bar {u}} \mid^2 )
+ \frac {1}{4} \mid {\cal M}_{g\bar {d}\bar {d} \to g\bar {d}\bar {d}} \mid^2 ]
         \nonumber    \\
& & \left. \times [f_{g1}f_{q2}f_{q3}(1+f_{g4})(1-f_{q5})(1-f_{q6})
           -f_{g4}f_{q5}f_{q6}(1+f_{g1})(1-f_{q2})(1-f_{q3})] \right\} ,     
         \nonumber    \\
\end{eqnarray}
and the transport equation for up-quark matter is
\begin{eqnarray}
\frac {\partial f_{q1}}{\partial t}
& + & \vec {\rm v}_1 \cdot \vec {\nabla}_{\vec {r}} f_{q1}
= -\frac {1}{2E_1} \int \frac {d^3p_2}{(2\pi)^32E_2}
\frac {d^3p_3}{(2\pi)^32E_3} \frac {d^3p_4}{(2\pi)^32E_4}
(2\pi)^4 \delta^4(p_1+p_2-p_3-p_4)
         \nonumber    \\
& & \times \left\{ g_G \mid {\cal M}_{ug \to ug} \mid^2
[f_{q1}f_{g2}(1-f_{q3})(1+f_{g4})-f_{q3}f_{g4}(1-f_{q1})(1+f_{g2})]  \right.
         \nonumber    \\
& & + g_Q (\frac {1}{2} \mid {\cal M}_{uu \to uu} \mid^2
+ \mid {\cal M}_{ud \to ud} \mid^2 
+ \mid {\cal M}_{u\bar {u} \to u\bar {u}} \mid^2
+ \mid {\cal M}_{u\bar {d} \to u\bar {d}} \mid^2 )
         \nonumber    \\
& & \times \left. [f_{q1}f_{q2}(1-f_{q3})(1-f_{q4})
-f_{q3}f_{q4}(1-f_{q1})(1-f_{q2})]   \right\}
         \nonumber    \\
& & -\frac {1}{2E_1} \int \frac {d^3p_2}{(2\pi)^32E_2}
\frac {d^3p_3}{(2\pi)^32E_3} \frac {d^3p_4}{(2\pi)^32E_4}
\frac {d^3p_5}{(2\pi)^32E_5} \frac {d^3p_6}{(2\pi)^32E_6}
         \nonumber    \\
& & \times (2\pi)^4 \delta^4(p_1+p_2+p_3-p_4-p_5-p_6)
\left\{ \frac {g_G^2}{4} \mid {\cal M}_{ugg \to ugg} \mid^2  \right.
         \nonumber    \\
& & \times [f_{q1}f_{g2}f_{g3}(1-f_{q4})(1+f_{g5})(1+f_{g6})
-f_{q4}f_{g5}f_{g6}(1-f_{q1})(1+f_{g2})(1+f_{g3})]
          \nonumber     \\
& & +g_Qg_G ( \frac {1}{2} \mid {\cal M}_{uug \to uug} \mid^2
+\mid {\cal M}_{udg \to udg} \mid^2
+\mid {\cal M}_{u\bar {u}g \to u\bar {u}g} \mid^2
+\mid {\cal M}_{u\bar {d}g \to u\bar {d}g} \mid^2 )
         \nonumber    \\
& & \times [f_{q1}f_{q2}f_{g3}(1-f_{q4})(1-f_{q5})(1+f_{g6})
-f_{q4}f_{q5}f_{g6}(1-f_{q1})(1-f_{q2})(1+f_{g3})]
          \nonumber     \\
& & + g_Q^2 [\frac {1}{12} \mid {\cal M}_{uuu \to uuu} \mid^2
+\frac {1}{4} ( \mid {\cal M}_{uud \to uud} \mid^2
              + \mid {\cal M}_{udu \to udu} \mid^2 )
+\frac {1}{4} \mid {\cal M}_{udd \to udd} \mid^2
         \nonumber    \\
& & +\frac {1}{2} \mid {\cal M}_{uu\bar {u} \to uu\bar {u}} \mid^2
    +\frac {1}{2} \mid {\cal M}_{uu\bar {d} \to uu\bar {d}} \mid^2
            + \mid {\cal M}_{ud\bar {u} \to ud\bar {u}} \mid^2
            + \mid {\cal M}_{ud\bar {d} \to ud\bar {d}} \mid^2
         \nonumber        \\
& & +\frac {1}{4} \mid {\cal M}_{u\bar {u}\bar {u}
                             \to u\bar {u}\bar {u}} \mid^2
    +\frac {1}{2} ( \mid {\cal M}_{u\bar {u}\bar {d}
                               \to u\bar {u}\bar {d}} \mid^2
+ \mid {\cal M}_{u\bar {d}\bar {u} \to u\bar {d}\bar {u}} \mid^2 )
+ \frac {1}{4}
  \mid {\cal M}_{u\bar {d}\bar {d} \to u\bar {d}\bar {d}} \mid^2 ]
         \nonumber   \\
& & \times \left. [f_{q1}f_{q2}f_{q3}(1-f_{q4})(1-f_{q5})(1-f_{q6})
-f_{q4}f_{q5}f_{q6}(1-f_{q1})(1-f_{q2})(1-f_{q3})] \right\} ,
         \nonumber    \\
\end{eqnarray}
where the massless gluon or the massless up quark has the velocity vector
${\vec {\rm v}}_1$ and the position vector $\vec r$, and the 
spin-color degeneracy factors are $g_{\rm G}=16$ for the gluon and 
$g_{\rm Q}=6$ for the quark, respectively. For the elastic 2-to-2 scattering
$p_1$ and $p_2$ ($p_3$ and $p_4$) are the four-momenta of the two initial 
(final) partons. For the elastic 3-to-3 scattering $p_1$, $p_2$, and $p_3$
($p_4$, $p_5$, and $p_6$) are the four-momenta of the three initial (final)
partons. $E_i$ is the energy component of $p_i$. 
$f_{gi}$ ($f_{qi}$) is the gluon (quark) distribution
function with the variable $p_i$. Denote the distribution
functions for the up quark, the down quark, the up antiquark, and the
down antiquark by $f_{ui}$, $f_{di}$, $f_{{\bar u}i}$, and $f_{{\bar d}i}$,
respectively, and, as assumed, they are identical, i.e.,
$f_{ui}=f_{di}=f_{{\bar u}i}=f_{{\bar d}i}=f_{qi}$.
In the transport equation for up-quark matter $f_{q1}$ is $f_{u1}$.
For the term with ${\cal M}_{u\bar {d} \to u\bar {d}}$, $f_{q2}$, $f_{q3}$, and
$f_{q4}$ are $f_{\bar {d}2}$, $f_{u3}$, and $f_{\bar {d}4}$, respectively. For
the term with ${\cal M}_{ud\bar {u} \to ud\bar {u}}$, $f_{q2}$, $f_{q3}$,
$f_{q4}$, $f_{q5}$, and $f_{q6}$ are $f_{d2}$, $f_{\bar {u}3}$, $f_{u4}$,
$f_{d5}$, and $f_{\bar {u}6}$, respectively. For the term with
${\cal M}_{gu\bar {d} \to gu\bar {d}}$ in the transport equation for gluon 
matter, $f_{q2}$, $f_{q3}$,
$f_{q5}$, and $f_{q6}$ are $f_{u2}$, $f_{\bar {d}3}$, $f_{u5}$,
and $f_{\bar {d}6}$, respectively. Other terms in the transport equations can
be understood in the same way. The transport equation for down-quark matter is
obtained from the transport equation for up-quark matter by the replacement,
$u \leftrightarrow d$ and $\bar {u} \leftrightarrow \bar d$; the transport
equation for up-antiquark matter by the replacement, $u \leftrightarrow \bar u$
and $d \leftrightarrow \bar d$; the transport equation for down-antiquark
matter by the 
replacement, $u \leftrightarrow \bar d$ and $d \leftrightarrow \bar u$.

\vspace{0.5cm}
\leftline {\bf V. NUMERICAL RESULTS AND DISCUSSIONS}
\vspace{0.5cm}

Quark-gluon matter initially produced in high-energy nucleus-nucleus collisions
can be simulated by HIJING \cite{WG}. For
central Au-Au collisions at $\sqrt {s_{NN}}=200$ GeV parton distributions 
inside matter take the form \cite{LMW}
\begin{equation}
f(k_\bot,y,r,z,t)=\frac {1}{16\pi R_A^2}g(k_\bot,y)\frac
{e^{-(z-t\tanh y)^2/2\Delta_k^2}}{\sqrt {2\pi}\Delta_k},
\end{equation}
in which $R_A$, $k_\bot$, $y$, $t$, $z$, and $r$ are the gold nucleus radius, 
transverse momentum, rapidity, time, coordinate in the longitudinal
direction, and radius in the transverse direction, respectively. $\Delta_k$ and
$g(k_\bot,y)$ are written as
\begin{displaymath}
\Delta_k \approx \frac {2}{k_\bot \cosh y},~~~~~~~~
g(k_\bot,y)=\frac {(2\pi)^3}{k_\bot \cosh y} \frac {dN}{dyd^2k_\bot}.
\end{displaymath}
The gluon and the quark have different $dN/dyd^2k_\bot$.
The distribution given by Eq. (37) is anisotropic because it has different
dependence on $k_\bot$ and $\cosh y$. We have assumed in Sec. IV that the
quark distribution is symmetric in flavor and identical with the antiquark
distribution. Then, 1500 gluons, 250 quarks or 250 antiquarks of the up 
or down flavor are created from the distribution function
in Eq. (37) within $-0.3<z<0.3$ fm and $r<R_A$ by the
rejection method. This is an initial condition of the transport equations at
$t=0.2$ fm/$c$. The coupling constant in the squared amplitudes
is taken to be $\alpha_{\rm s}={\rm g}^2_{\rm s}/4\pi=0.5$ 
in solving the transport equations. The screening mass formulated in Refs.
\cite{BMW,BMS,KK} is calculated from the gluon and quark distribution
functions and is used to regularize propagators in the squared amplitudes for
both the elastic 2-to-2 scattering and the elastic 3-to-3 scattering.

The two thousand, five hundred partons are inhomogeneous in coordinate space
and anisotropic in momentum space, and scatter to have their momenta changed.
Scattering of two partons occurs when the two partons have the closest
distance less than the square root of the ratio of the cross section for 2-to-2
scattering to $\pi$. The cross section depends on the total energy of the two
colliding partons in their center-of-momentum system and can be found in Ref.
\cite{XX} for the elastic scattering of one gluon and one quark, one gluon
and one antiquark, or two gluons and in Ref. \cite{XMCW} 
for one quark and one antiquark, two quarks, or two antiquarks,
where the fraction 8/9 in Eq. (5) should be replaced
by 4/9. Scattering of three partons occurs if the three partons are in a sphere
of which the center is at the center-of-mass of the three partons and of which
the radius $r_{\rm hs}$ is \cite{XSCZ}
\begin{eqnarray}
\pi r_{\rm hs}^2
& = & \frac {1}{n_{\rm f}!} \int \frac {d^3p_4}{(2\pi)^32E_4}
\frac {d^3p_5}{(2\pi)^32E_5} \frac {d^3p_6}{(2\pi)^32E_6}
         \nonumber    \\
& &
\times (2\pi)^4 \delta^4(p_1+p_2+p_3-p_4-p_5-p_6)
\mid {\cal M}_{3 \to 3} \mid^2,
\end{eqnarray}
where ${\cal M}_{3 \to 3}$ is the amplitude for the 3-parton scattering,
and $n_{\rm f}$ is the number of identical partons in the final states. For
example, ${\cal M}_{3 \to 3}$ is ${\cal M}_{ud\bar {u} \to ud\bar u}$ for
$ud\bar {u} \to ud\bar u$ or ${\cal M}_{ggu \to ggu}$ for $ggu \to ggu$;
$n_{\rm f}$ equals 2 for $ggu \to ggu$, $ggd \to ggd$, 
$gg\bar {u} \to gg\bar u$, $gg\bar {d} \to gg\bar d$, $guu \to guu$, 
$gdd \to gdd$,  $g\bar {u}\bar {u} \to g\bar {u}\bar {u}$,
$g\bar {d}\bar {d} \to g\bar {d}\bar {d}$,
$u\bar {u}\bar {u} \to u\bar {u}\bar {u}$, or
$u\bar {d}\bar {d} \to u\bar {d}\bar {d}$.

Without scattering every parton moves in a straight line and the anisotropy of
quark-gluon matter indicated by Eq. (37)
remains. Due to scattering, partons' momenta are changed, and
the anisotropy is removed. For example, at $t=0.2$ fm/$c$
the distribution of gluons in the transverse direction is very
different from that in the longitudinal direction.
But such difference disappears after
scattering in a short period of time. This is because the scattering alters the
momentum distribution functions in the two directions so that both
become identical at a moment. Hence, we observe gluon momentum
distribution functions in the three directions marked by the three angles, 
$0^{\rm o}$, $45^{\rm o}$, and $90^{\rm o}$, 
relative to an incoming beam direction, and show them by the dotted, dashed and
dot-dashed curves in Fig. 7 at a time of the order of 0.52 fm/$c$. 
The curves overlapped are
fitted to the J\"uttner distribution shown by the solid curve,
\begin{equation}
f_g(\vec {p})=\frac {\lambda_g}{{\rm e}^{\mid \vec {p}\mid/T}-\lambda_g},
\end{equation}
with the temperature $T=0.52$ GeV and the fugacity $\lambda_g =0.328$. Since
such a thermal state of gluon matter with the temperature is established at 
$t=0.52$ fm/$c$, the thermalization time of gluon matter is 0.32 fm/$c$.
Nevertheless, we have to mention that, while the thermal state of gluon matter
is established, quark matter has not been in a thermal state. 
This means that quark matter does not thermalize as quick as gluon matter. 
We find that the quark momentum distribution functions in the three directions
first overlap at a time of the order of 0.86 fm/$c$, 
and they are shown in Fig. 8. Then, the thermalization time of quark
matter is 0.66 fm/$c$. The overlapped curves are also fitted to the J\"uttner
distribution,
\begin{equation}
f_q(\vec {p})=\frac {\lambda_q}{{\rm e}^{\mid \vec {p}\mid/T}+\lambda_q},
\end{equation}
with $T=0.46$ GeV and the fugacity $\lambda_q=0.143$.

While the elastic gluon-gluon-quark scattering and the elastic
gluon-gluon-antiquark scattering are absent, a thermalization time of the
order of 0.48 fm/$c$ was obtained for gluon matter in Ref. \cite{XX}. 
In the present work the elastic scattering of $ggq$ and $gg\bar q$ shortens
thermalization time of gluon matter to
0.32 fm/$c$. In finding the solutions of the transport equations, we
know that most partons need to scatter one or two times. At least,
partons should scatter one time so that the thermalization time has a lower
limit and can not be zero.

In Ref. \cite{XMCW} thermalization of quark matter and antiquark matter was
studied with the elastic scattering of quark-quark, antiquark-antiquark,
quark-antiquark, quark-quark-quark, antiquark-antiquark-antiquark,
quark-quark-antiquark, and quark-antiquark-an- tiquark. 
The thermalization times of quark matter and antiquark matter were 
of the order of 1.55 fm/$c$. Furthermore,
in Ref. \cite{XX} a thermalization time of the order of 1.36 fm/$c$ was
obtained for quark matter while the elastic scattering of gluon-quark-quark,
gluon-antiquark-antiquark, and gluon-quark-antiquark is included, and the
thermalization time was shortened by 0.19 fm /$c$.
Relative to this thermalization time 1.36 fm/$c$, the thermalization time 0.66
fm/$c$ in the present work means that the elastic scattering of $ggq$ and
$gg\bar q$ shortens the thermalization time by the amount 0.7 fm/$c$. 
Compared to 0.19 fm/$c$, the amount is large. This is due to two facts. 
One is that
the gluon number density is 6 times the quark number density in a flavor;
another is that the squared amplitude for the elastic gluon-gluon-quark
scattering is an order of magnitude larger than that for the elastic
gluon-quark-quark scattering.

The thermalization times of gluon matter and quark matter are 0.32 fm/$c$ and
0.66 fm/$c$, respectively, which indicate early thermalization of both gluon
matter and quark matter. Even though the thermalization times of gluon matter
and quark matter are both less than 1 fm/$c$, they are different. The elastic
2-gluon scattering and the elastic 3-gluon scattering directly cause a change
of gluon 
distribution and indirectly cause changes of quark distribution and antiquark
distribution by the elastic scattering of gluon-quark, gluon-antiquark,
gluon-quark-quark, gluon-quark-antiquark, gluon-antiquark-antiquark,
gluon-gluon-quark, and gluon-gluon-antiquark. The elastic two-body scattering
and the elastic three-body scattering inside quark and antiquark matter
directly cause changes of quark distribution
and antiquark distribution and indirectly cause a change of gluon
distribution by the elastic scattering of gluon-quark, gluon-antiquark,
gluon-quark-quark, gluon-quark-antiquark, gluon-antiquark-antiquark, 
gluon-gluon-quark, and gluon-gluon-antiquark. In Ref. \cite{XX} we have shown
that the elastic scattering of gluon-quark, gluon-antiquark, gluon-quark-quark,
gluon-quark-antiquark, and gluon-antiquark-antiquark gi- ves similar
contributions to the variation of the gluon distribution function and the
variation of the quark (antiquark) distribution function. In the present work
$\frac {g_Gg_Q}{2} ( \mid {\cal M}_{ggu \to ggu} \mid^2
+\mid {\cal M}_{ggd \to ggd} \mid^2
+\mid {\cal M}_{gg\bar {u} \to gg\bar {u}} \mid^2 
+\mid {\cal M}_{gg\bar {d} \to gg\bar {d}} \mid^2 )$ in Eq. (35) is three times
$\frac {g_G^2}{4} \mid {\cal M}_{ugg \to ugg} \mid^2$ in Eq. (36). The gluon
distribution function $f_{gi}$ is roughly twice the quark distribution function
$f_{qi}$. Then, the variation of the gluon distribution function is roughly
three times the variation of the quark (antiquark) distribution function. 
Now we can understand
the difference of the thermalization times of gluon matter and quark matter
from the facts: gluon matter is denser than quark matter; the elastic
gluon-gluon (gluon-gluon-gluon) scattering has a larger squared amplitude than
the elastic quark-quark or quark-antiquark (quark-quark-quark or 
quark-quark-antiquark) scattering; the contribution of the elastic
gluon-gluon-quark scattering to the variation of the gluon distribution
function is larger than the contribution to the variation of the quark
distribution function.

Let us consider one unrealistic case, but it can help us better understand
the difference between the thermalization time of gluon matter and 
that of quark matter.
The case is that the quark number equals the gluon number 1500. Then, the
thermalization time of quark matter is reduced by the increase of quark number
from 500 to 1500, and the quark distribution function $f_{qi}$ is roughly 
the gluon distribution function $f_{gi}$. The difference between the variation
of the gluon distribution and the variation of the quark distribution is 
mainly determined by the spin-color degeneracy factors and the squared 
amplitudes as shown in Eqs. (35) and (36). 
We calculate the squared amplitudes for various types of elastic
scattering and obtain the following results:
$g_Q ( \mid {\cal M}_{gu \to gu} \mid^2 + \mid {\cal M}_{gd \to gd} \mid^2
+ \mid {\cal M}_{g\bar {u} \to g\bar {u}} \mid^2
+ \mid {\cal M}_{g\bar {d} \to g\bar {d}} \mid^2 )$ in Eq. (35) is near
$g_G \mid {\cal M}_{ug \to ug} \mid^2$ in Eq. (36);
$g_Q^2 [\frac {1}{4} \mid {\cal M}_{guu \to guu} \mid^2
+\frac {1}{2} ( \mid {\cal M}_{gud \to gud} \mid^2
              + \mid {\cal M}_{gdu \to gdu} \mid^2 )
+\frac {1}{4} \mid {\cal M}_{gdd \to gdd} \mid^2
+ \mid {\cal M}_{gu\bar {u} \to gu\bar {u}} \mid^2
+ \mid {\cal M}_{gu\bar {d} \to gu\bar {d}} \mid^2
+ \mid {\cal M}_{gd\bar {u} \to gd\bar {u}} \mid^2
+ \mid {\cal M}_{gd\bar {d} \to gd\bar {d}} \mid^2
+\frac {1}{4} \mid {\cal M}_{g\bar {u}\bar {u} \to g\bar {u}\bar {u}} \mid^2
+\frac {1}{2} ( \mid {\cal M}_{g\bar {u}\bar {d} \to g\bar {u}\bar {d}} \mid^2 
+\mid {\cal M}_{g\bar {d}\bar {u} \to g\bar {d}\bar {u}} \mid^2 )
+\frac {1}{4} \mid {\cal M}_{g\bar {d}\bar {d} \to g\bar {d}\bar {d}} \mid^2 ]$
is near
$g_Qg_G ( \frac {1}{2} \mid {\cal M}_{uug \to uug} \mid^2
+\mid {\cal M}_{udg \to udg} \mid^2
+\mid {\cal M}_{u\bar {u}g \to u\bar {u}g} \mid^2
+\mid {\cal M}_{u\bar {d}g \to u\bar {d}g} \mid^2 )$;
$\frac {g_Gg_Q}{2} ( \mid {\cal M}_{ggu \to ggu} \mid^2
+\mid {\cal M}_{ggd \to ggd} \mid^2
+\mid {\cal M}_{gg\bar {u} \to gg\bar {u}} \mid^2 
+\mid {\cal M}_{gg\bar {d} \to gg\bar {d}} \mid^2 )$ is three times
$\frac {g_G^2}{4} \mid {\cal M}_{ugg \to ugg} \mid^2$;
$\frac {g_G}{2} \mid {\cal M}_{gg \to gg} \mid^2 $ is over three times
$g_Q (\frac {1}{2} \mid {\cal M}_{uu \to uu} \mid^2
+ \mid {\cal M}_{ud \to ud} \mid^2 
+ \mid {\cal M}_{u\bar {u} \to u\bar {u}} \mid^2
+ \mid {\cal M}_{u\bar {d} \to u\bar {d}} \mid^2 )$;
$\frac {g_G^2}{12} \mid {\cal M}_{ggg \to ggg} \mid^2 $ is an order of
magnitude larger than
$g_Q^2 [\frac {1}{12} \mid {\cal M}_{uuu \to uuu} \mid^2
+\frac {1}{4} ( \mid {\cal M}_{uud \to uud} \mid^2
              + \mid {\cal M}_{udu \to udu} \mid^2 )
+\frac {1}{4} \mid {\cal M}_{udd \to udd} \mid^2
+\frac {1}{2} \mid {\cal M}_{uu\bar {u} \to uu\bar {u}} \mid^2
+\frac {1}{2} \mid {\cal M}_{uu\bar {d} \to uu\bar {d}} \mid^2
            + \mid {\cal M}_{ud\bar {u} \to ud\bar {u}} \mid^2
            + \mid {\cal M}_{ud\bar {d} \to ud\bar {d}} \mid^2
+\frac {1}{4} \mid {\cal M}_{u\bar {u}\bar {u} \to u\bar {u}\bar {u}} \mid^2
+\frac {1}{2} ( \mid {\cal M}_{u\bar {u}\bar {d} \to u\bar {u}\bar {d}} \mid^2
+\mid {\cal M}_{u\bar {d}\bar {u} \to u\bar {d}\bar {u}} \mid^2 )
+\frac {1}{4} 
\mid {\cal M}_{u\bar {d}\bar {d} \to u\bar {d}\bar {d}} \mid^2 ]$.
Hence, the difference between the variation of the gluon distribution
and the variation of the quark distribution is mainly
caused by the elastic scattering of gluon-gluon, gluon-gluon-gluon,
quark-quark, quark-antiquark, quark-quark-quark,
quark-quark-antiquark, quark-antiquark-antiquark, gluon-gluon-quark, and
gluon-gluon-antiquark, and the difference between
the thermalization time of gluon matter and that of quark matter certainly
exists. In the case that the
number of quarks equals the number of gluons, the thermalization time of quark
matter approaches that of gluon matter, but quark matter still can not
thermalize faster than gluon matter.

The gluon multiplication process $gg \to ggg$ was used by Bir$\acute {\rm o}$
et al. to study chemical equilibration of quark-gluon plasmas 
\cite{BDMTW}, and such a
process also contributes to the thermalization of quark-gluon matter
\cite{SM,XG}. In Ref. \cite{LMW} the Landau-Pomeranchuk-Migdal (LPM) effect is
taken into account to suppress the gluon radiation in $gg \to ggg$. The LPM
effect appears while the radiation formation time is long compared to the mean
free path. The effective formation time in QCD is \cite{WGP}
\begin{equation}
\tau_{\rm QCD}=\frac {C_A}{2C_2} \frac {2\cosh y_g}{k_{g\bot}},
\end{equation}
where $C_2=C_A=3$ for a gluon, and $k_{g\bot}$ and $y_g$ are the transverse
momentum and the rapidity of the radiated gluon, respectively. The larger the
absolute value of the rapidity, or the smaller the transverse momentum,
the longer the effective formation time. At $y_g=0$, $\tau_{\rm QCD}$ equal to
the mean free path $\bar \lambda$ gives the minimum of the transverse momentum,
\begin{equation}
k_{g\bot\rm min}=\frac {1}{\bar \lambda},
\end{equation}
over which the LPM effect turns up. The mean free path is given by
\begin{equation}
\bar {\lambda}=\frac {1}{\sqrt {2}n\sigma_2},
\end{equation}
where $n$ is the quark-gluon number density and $\sigma_2$ is the cross section
for two-parton scattering. While $n$ or $\sigma_2$ increases, the mean free 
path decreases. We assume that $\sigma_2$ is 3 mb, which is
smaller than the value 5 mb in Ref. \cite{XG}. Quark-gluon matter is produced
at the time 0.2 fm/$c$, 
and gluon matter and quark matter arrive at their thermal states at 0.52 
fm/$c$ and 0.86 fm/$c$, respectively. In Table I we list the number density,
the mean free path, and the minimum of the transverse momentum
at the three times. $k_{g\bot\rm min}$ is 2.7 GeV/$c$ at $t=0.2$ fm/$c$, and
the gluon radiation is highly suppressed. From $t=0.2$ fm/$c$ to 0.86 fm/$c$
the minimum of the transverse momentum decreases, and $gg \to ggg$ is gradually
involved in thermalization. Thermalization mainly relates to
the distribution functions in the momentum region $\mid \vec {p} \mid < 1.5$ 
GeV/$c$. $k_{g\bot\rm min}$ is 1.42 GeV/$c$ at 0.52 fm/$c$. The
influence of $gg \to ggg$ on the gluon distribution for 
$\mid \vec {p} \mid < 1.5$ GeV/$c$ should be very limited,
so should the influence on the early thermalization of gluon matter. 
$k_{g\bot\rm min}$ is 0.91 GeV/$c$ at 0.86 fm/$c$.
Since the radiated gluon will scatter with quarks
inside quark-gluon matter, the quark distribution and the early
thermalization of quark matter are expected to be moderately affected.
Therefore, the gluon multiplication process may slightly modify
the thermalization time of gluon matter and may moderately modify the
thermalization time of quark matter. The absence of the process is reliable in
the early thermalization of gluon matter but approximate in the early
thermalization of quark matter.

\vspace{0.5cm}
\leftline{\bf VI. SUMMARY}
\vspace{0.5cm}

We have derived the squared amplitude for the elastic gluon-gluon-quark
scattering in perturbative QCD. One hundred and twenty-three Feynman diagrams
at the tree level are introduced, and a method is presented to derive the
individually squared amplitudes and the interference terms. 
With the squared amplitudes for the elastic $ggq$ scattering and the
elastic $gg\bar q$ scattering, the transport equations for gluons,
up quarks, down quarks, up antiquarks and down antiquarks are solved to get
thermal states and
thermalization times of gluon matter, quark matter and antiquark matter which
are initially produced in central Au-Au collisions at the highest RHIC energy
and of which the evolution is governed by elastic 2-to-2 scattering and elastic
3-to-3 scattering.
Early thermalization of initially produced quark-gluon matter is indicated by
the short thermalization times of gluon matter and quark matter. 
However, the thermalization time of gluon matter
differs from that of quark matter. So far, we have answered
how and when initially produced quark-gluon matter establishes a thermal state.

\vspace{0.5cm}
\leftline{\bf ACKNOWLEDGMENT}
\vspace{0.5cm}
This work was supported by the National Natural Science Foundation of China
under Grant No. 11175111.

\newpage
\begin{figure}[htbp]
  \centering
    \includegraphics[scale=0.8]{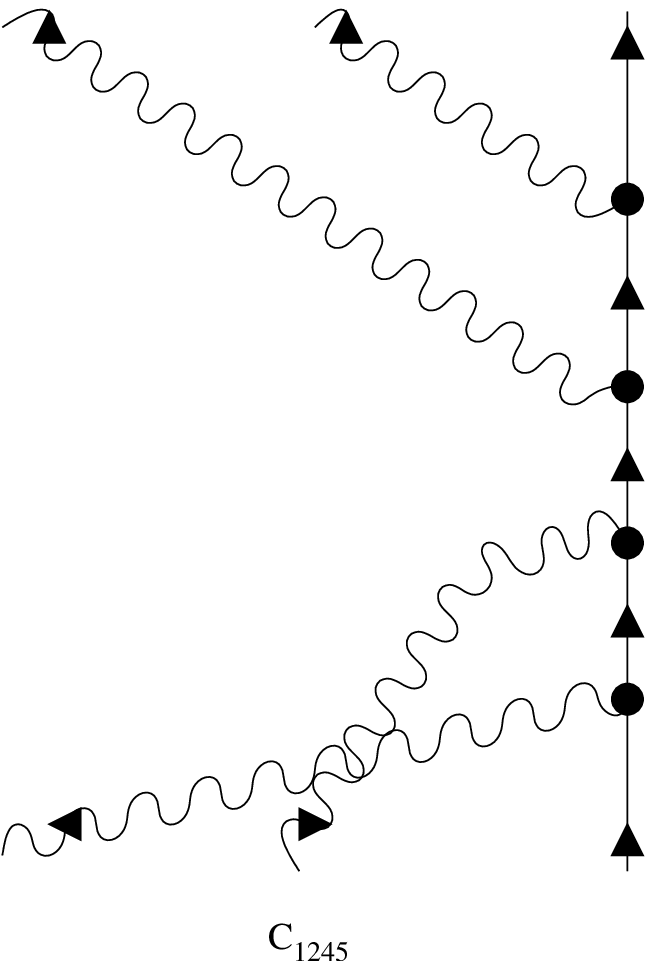}%
\caption{Elastic gluon-gluon-quark scattering.}
\label{fig1}
\end{figure}

\begin{figure}[htbp]
  \centering
    \includegraphics[width=42mm,height=65mm,angle=0]{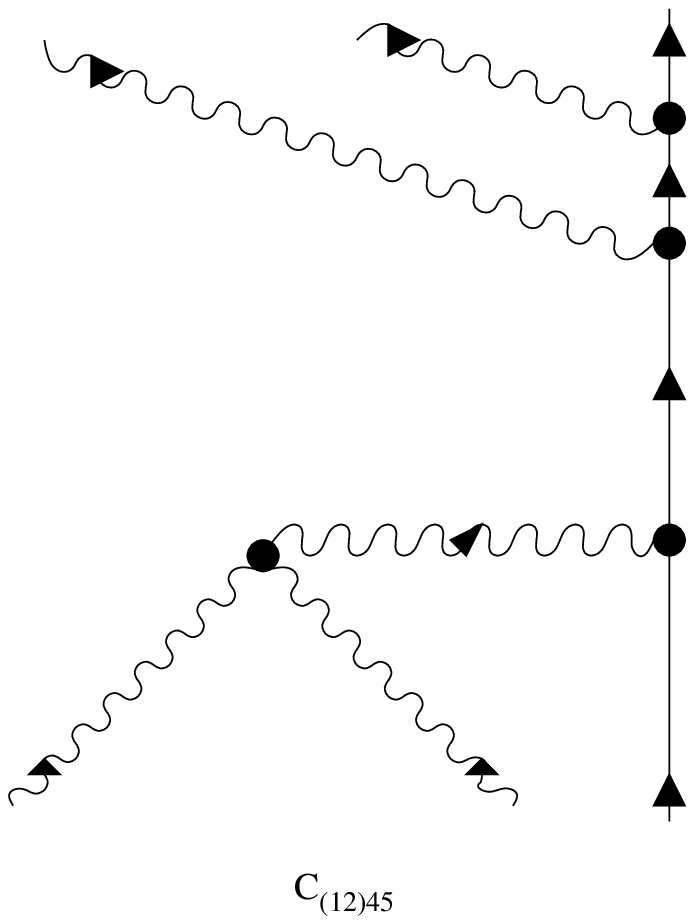}
      \hspace{1.2cm}
    \includegraphics[width=42mm,height=65mm,angle=0]{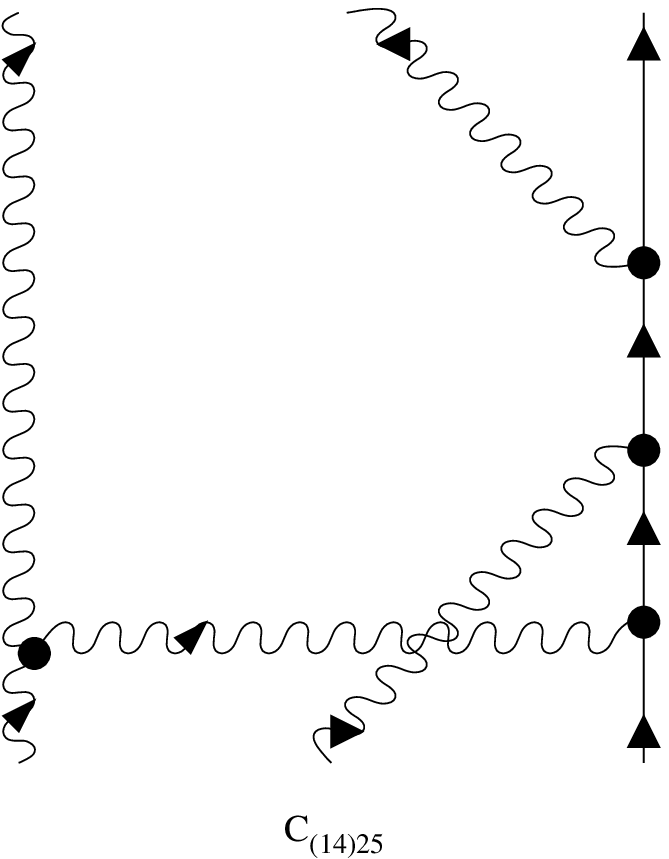}
      \hspace{1.2cm}
    \includegraphics[width=42mm,height=65mm,angle=0]{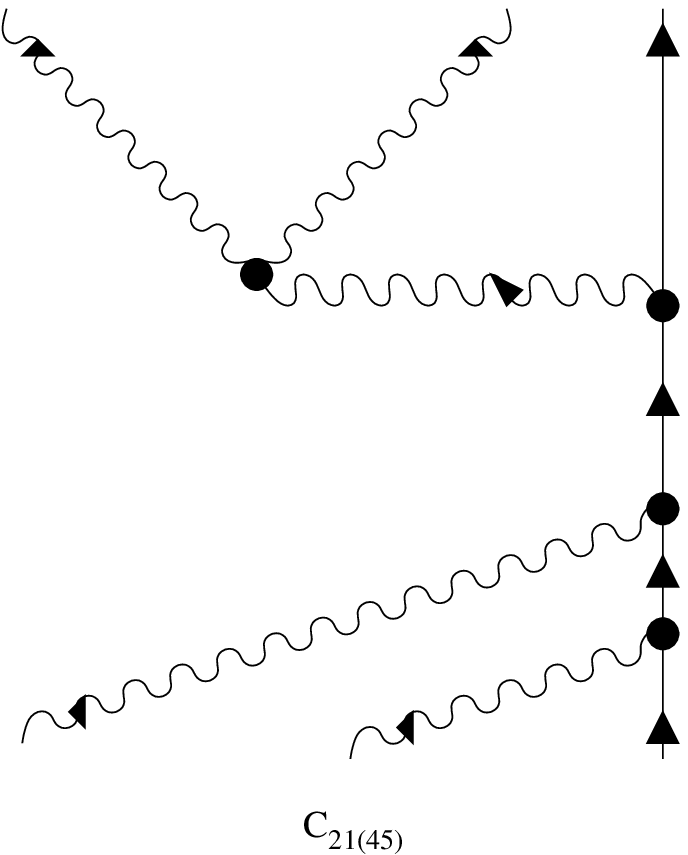}
\caption{Elastic gluon-gluon-quark scattering.}
\label{fig2}
\end{figure}

\newpage
\begin{figure}
  \centering
    \includegraphics[width=42mm,height=65mm,angle=0]{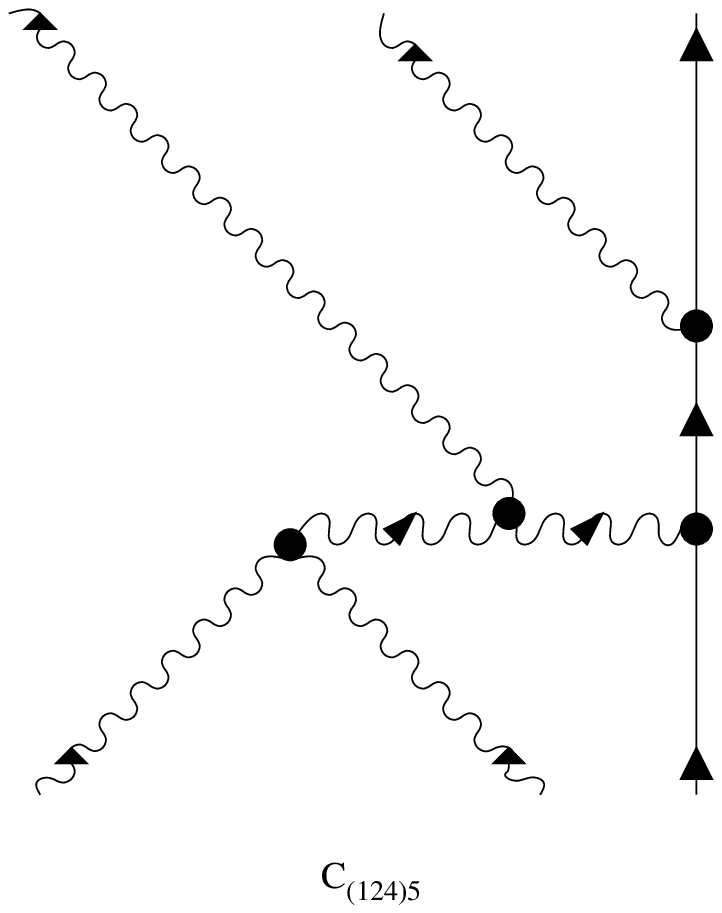}
      \hspace{1.2cm}
    \includegraphics[width=42mm,height=65mm,angle=0]{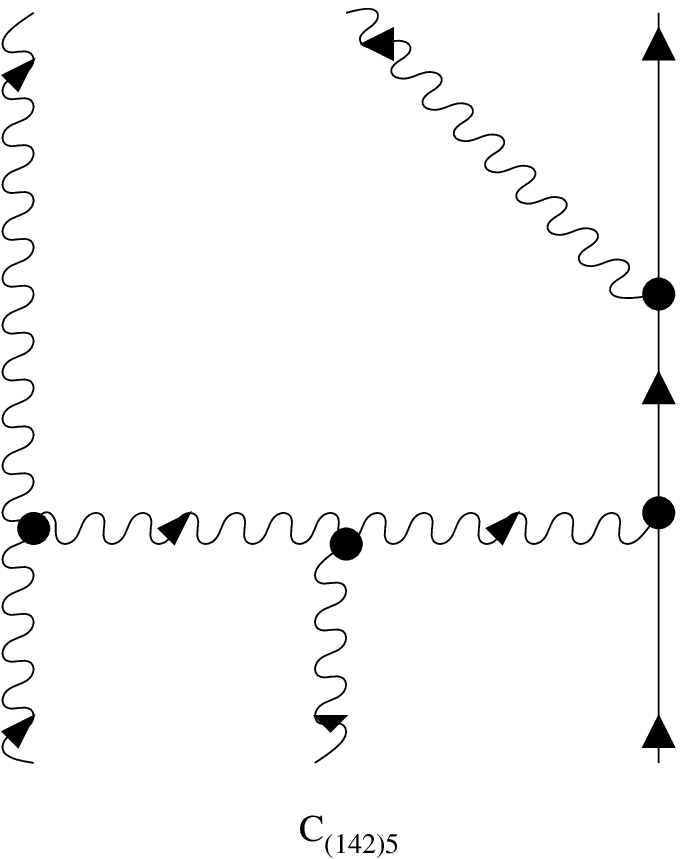}
      \hspace{1.2cm}
    \includegraphics[width=42mm,height=65mm,angle=0]{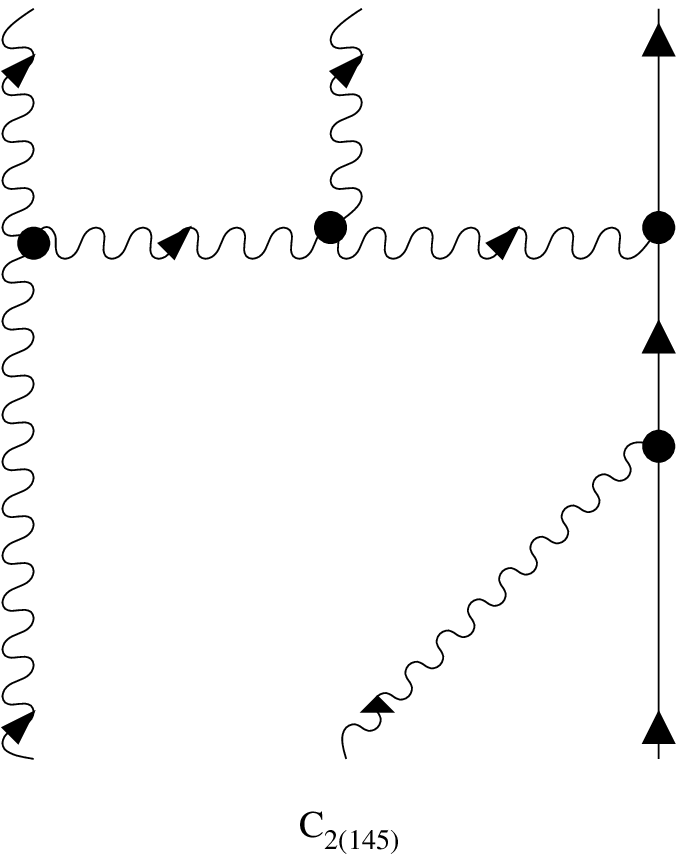}
      \vskip 26pt
    \includegraphics[width=42mm,height=65mm,angle=0]{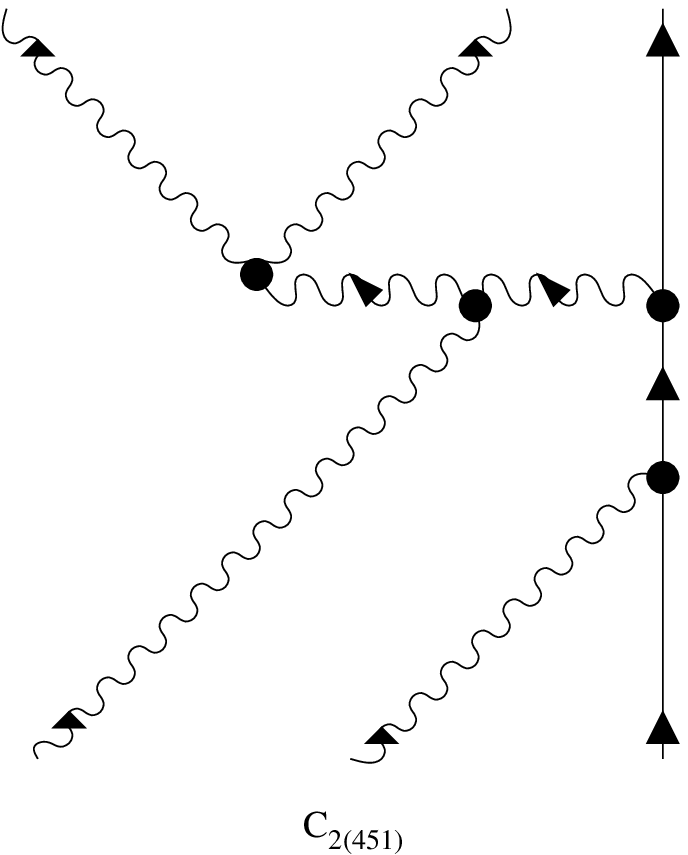}
      \hspace{1.2cm}
    \includegraphics[width=42mm,height=65mm,angle=0]{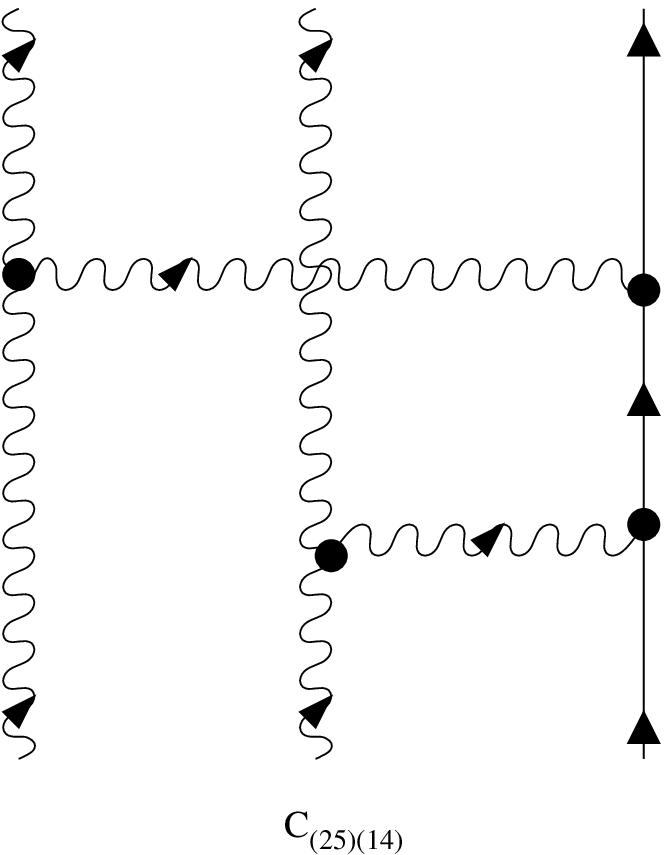}
      \hspace{1.2cm}
    \includegraphics[width=42mm,height=65mm,angle=0]{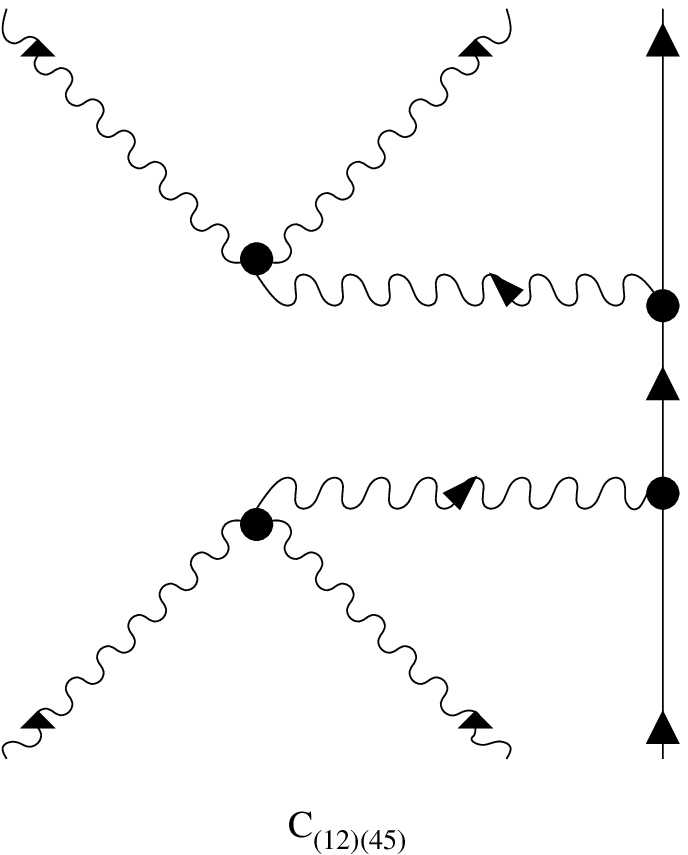}
\caption{Elastic gluon-gluon-quark scattering.}
\label{fig3}
\end{figure}

\newpage
\begin{figure}
  \centering
    \includegraphics[width=42mm,height=65mm,angle=0]{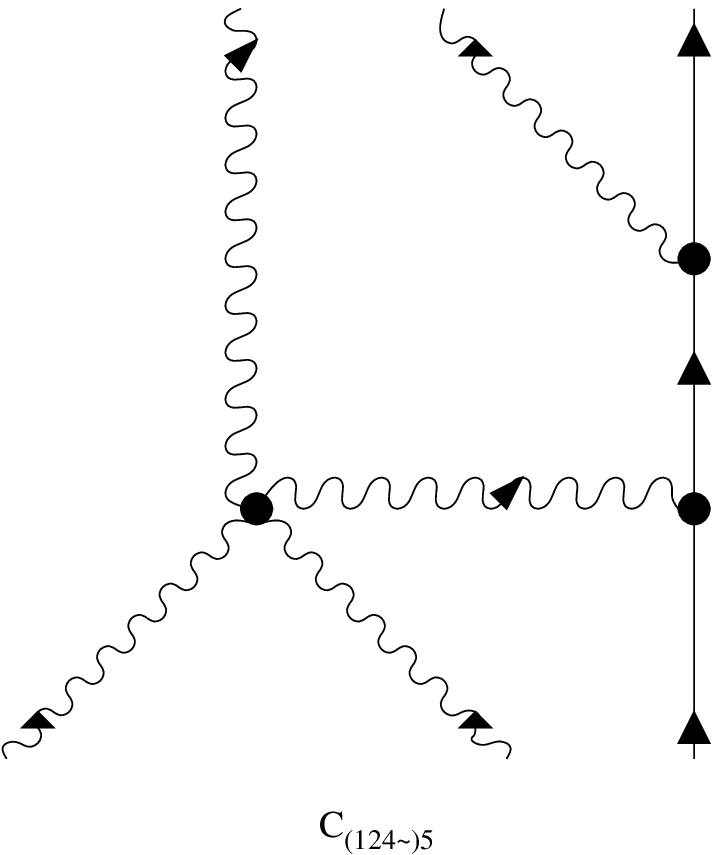}
      \hspace{1.2cm}
    \includegraphics[width=42mm,height=65mm,angle=0]{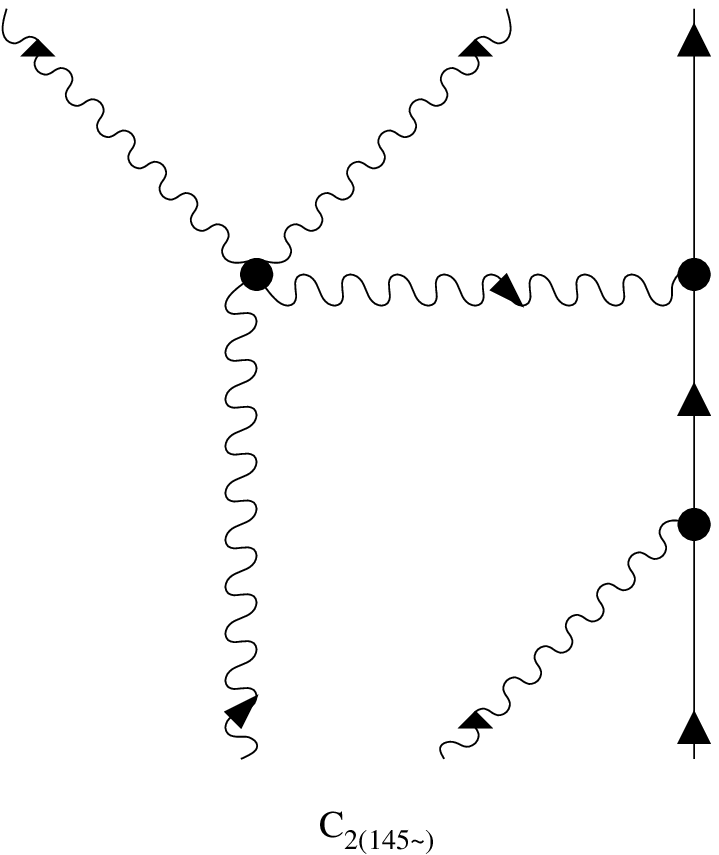}
\caption{Elastic gluon-gluon-quark scattering.}
\label{fig4}
\end{figure}

\newpage
\begin{figure}
  \centering
    \includegraphics[width=42mm,height=65mm,angle=0]{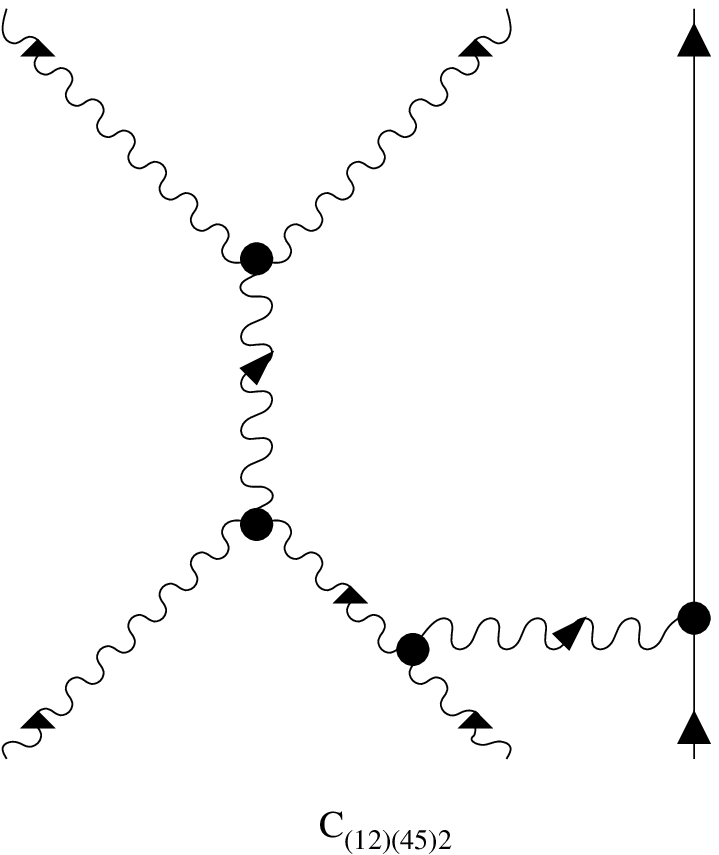}
      \hspace{1.2cm}
    \includegraphics[width=42mm,height=65mm,angle=0]{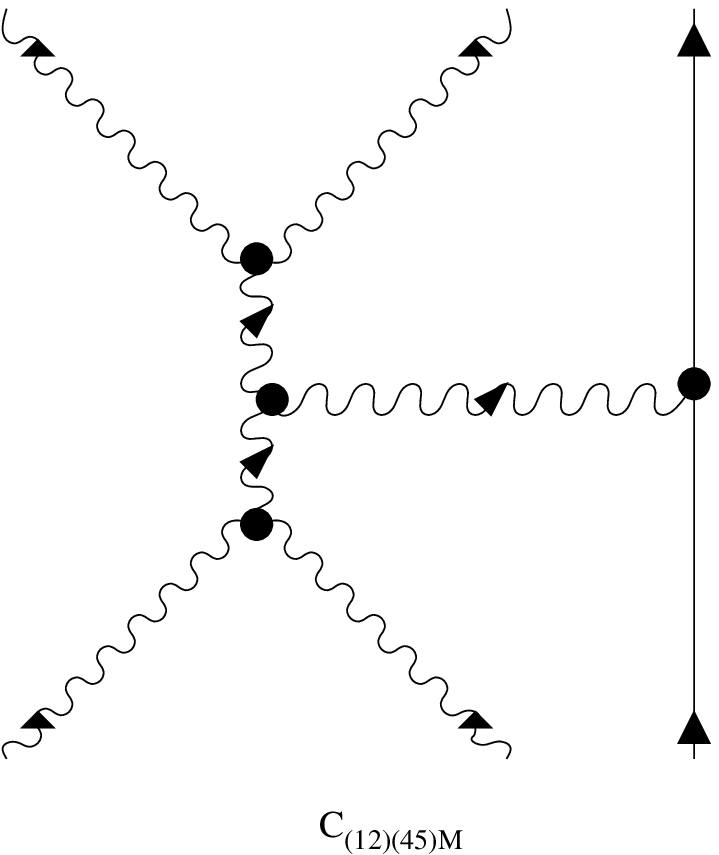}
      \vskip 26pt
    \includegraphics[width=42mm,height=65mm,angle=0]{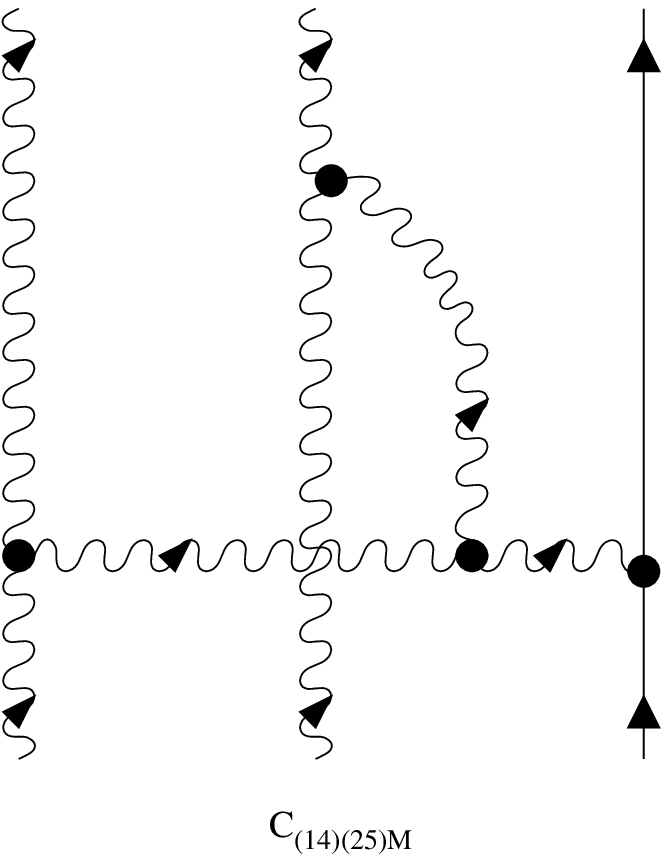}
      \hspace{1.2cm}
    \includegraphics[width=42mm,height=65mm,angle=0]{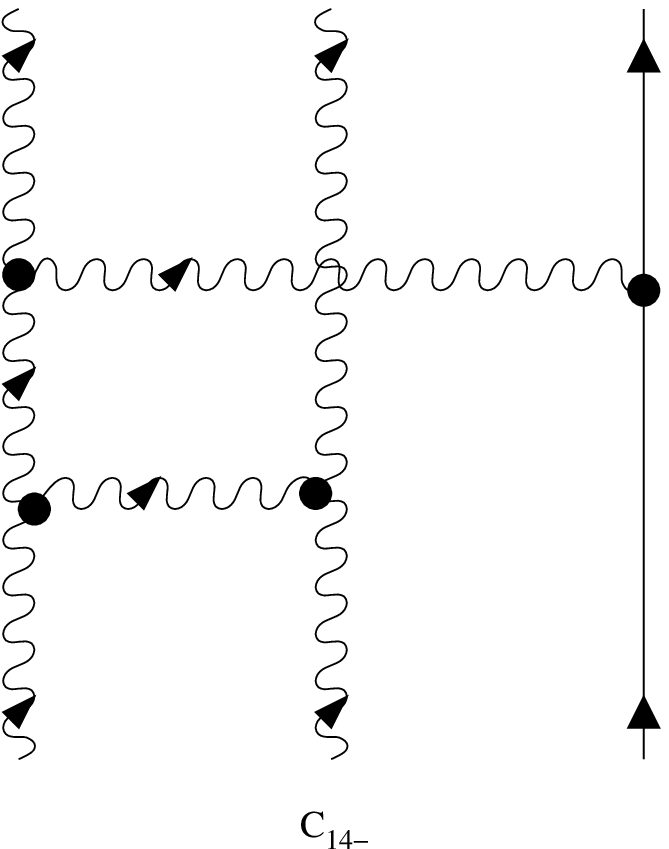}
\caption{Elastic gluon-gluon-quark scattering.}
\label{fig5}
\vspace{3cm}
\end{figure}

\newpage
\begin{figure}
  \centering
    \includegraphics[width=42mm,height=65mm,angle=0]{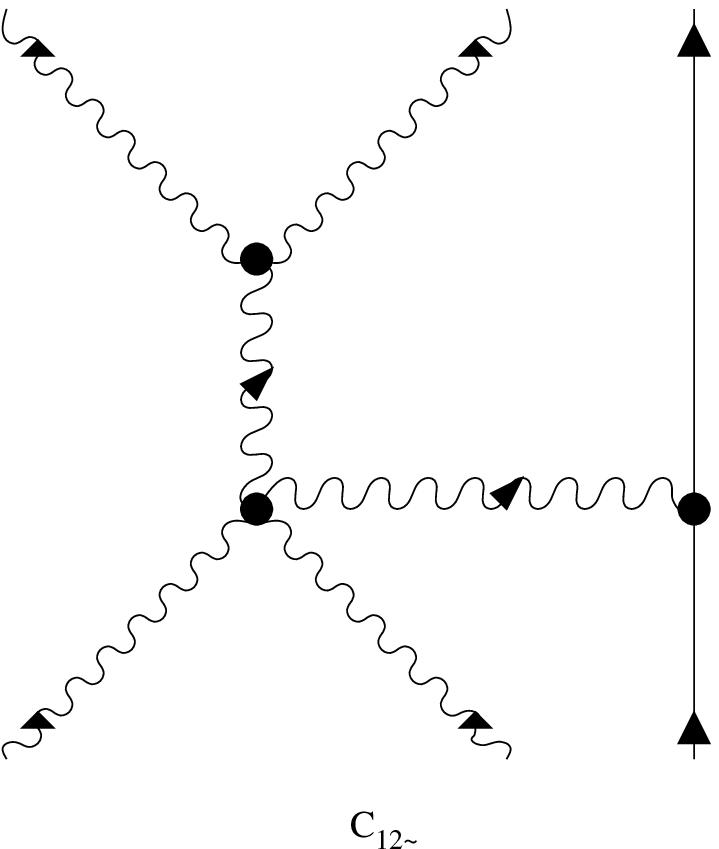}
      \hspace{1.2cm}
    \includegraphics[width=42mm,height=65mm,angle=0]{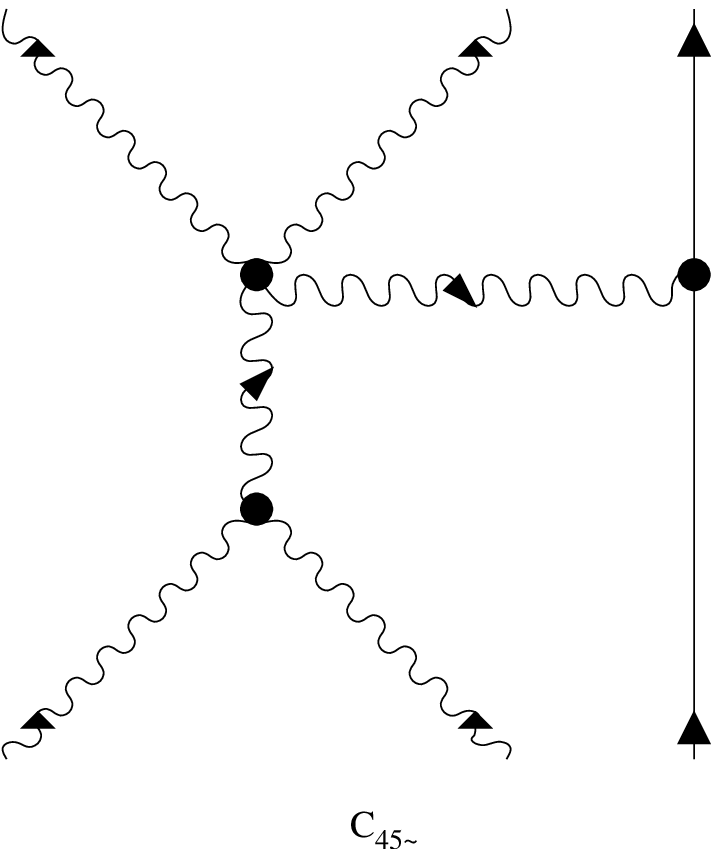}
      \vskip 26pt
    \includegraphics[width=42mm,height=65mm,angle=0]{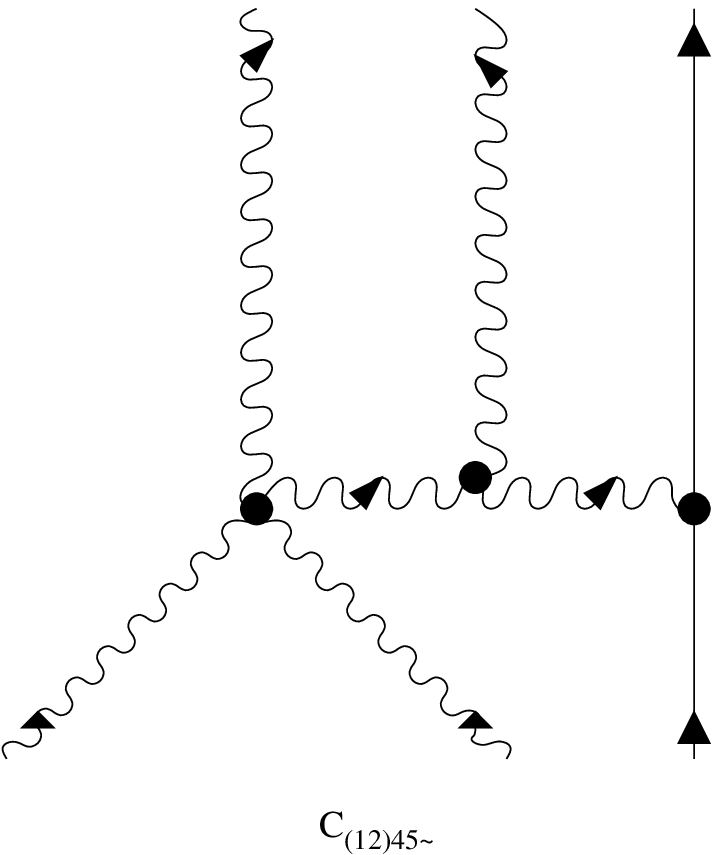}
      \hspace{1.2cm}
    \includegraphics[width=42mm,height=65mm,angle=0]{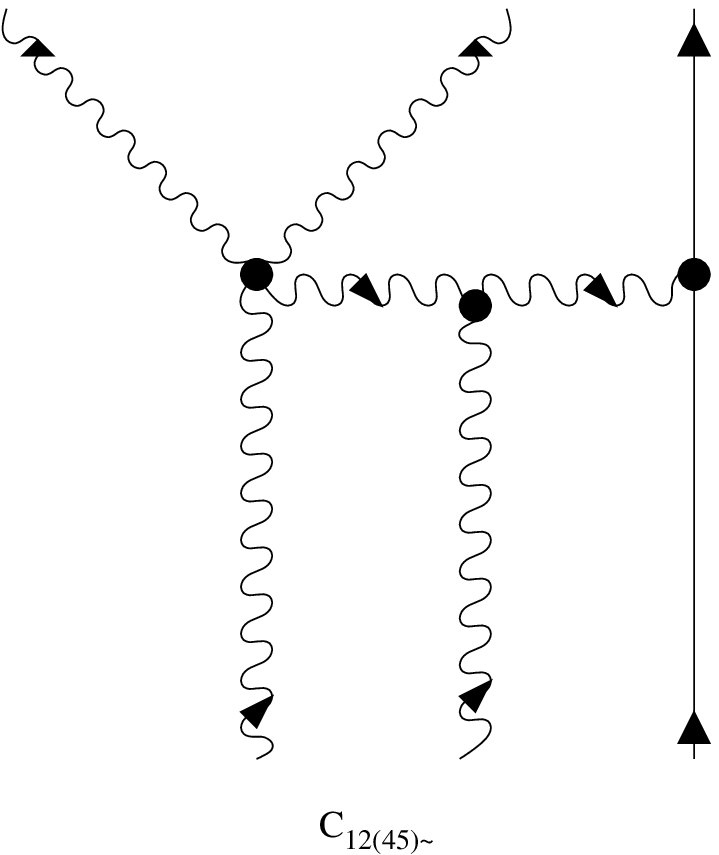}
      \hspace{1.2cm}
    \includegraphics[width=42mm,height=65mm,angle=0]{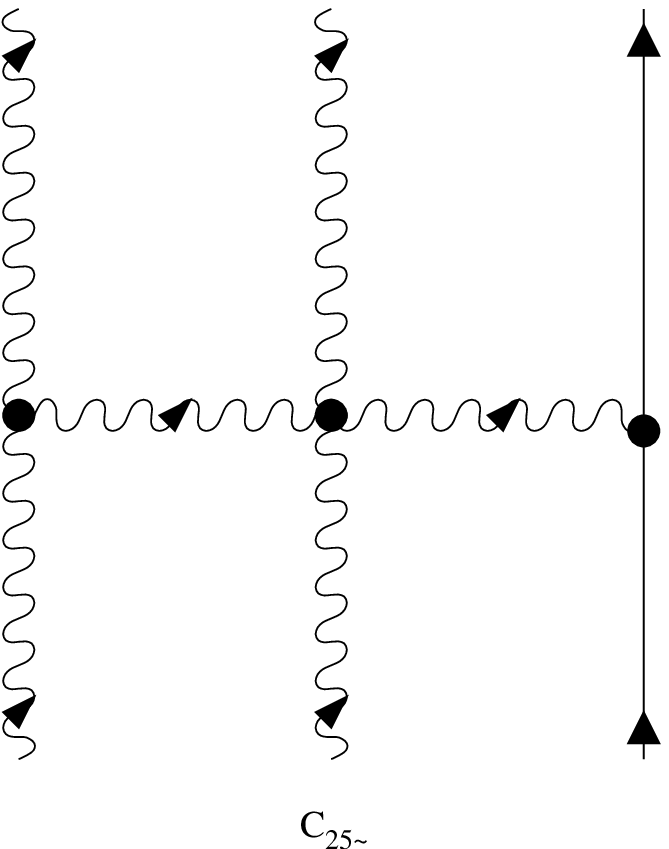}
\caption{Elastic gluon-gluon-quark scattering.}
\label{fig6}
\vspace{3cm}
\end{figure}

\newpage
\begin{figure}
  \centering
    \includegraphics[width=120mm,height=80mm,angle=0]{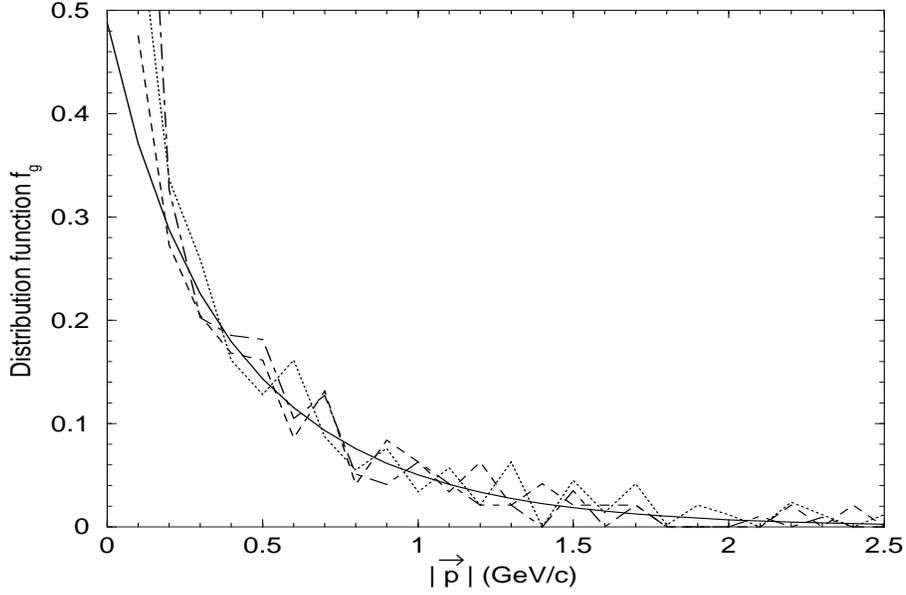}
\caption{Gluon distribution functions versus momentum in different directions
while gluon matter arrives at a thermal state. The dotted, dashed and 
dot-dashed curves correspond to the angles, $0^{\rm o}, 45^{\rm o}$, and 
$90^{\rm o}$, relative to one incoming beam direction, respectively.
The solid curve represents the thermal distribution function. The 
thermalization time of gluon matter is 0.32 fm/$c$.}
\label{fig7}
\end{figure}

\newpage
\begin{figure}
  \centering
    \includegraphics[width=120mm,height=80mm,angle=0]{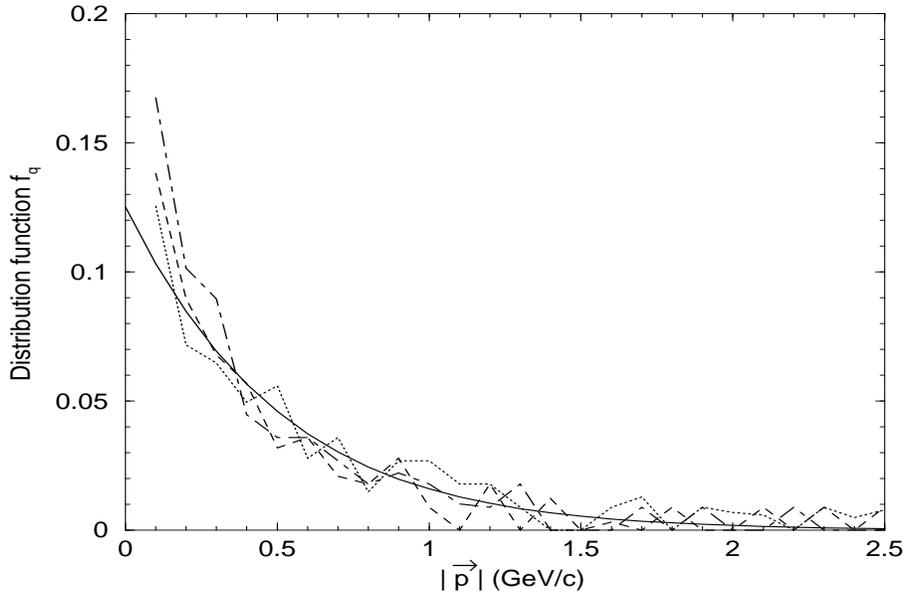}
\caption{Same as Fig. 7, except for quark distribution functions 
while quark matter arrives at a thermal state. The thermalization time of quark
matter is 0.66 fm/$c$.}
\label{fig8}
\end{figure}

\newpage
\begin{table}
\centering \caption{$n$, $\bar \lambda$, and $k_{g\bot \rm min}$ 
at the three times
which correspond to the formation of quark-gluon matter, the thermal state of
gluon matter, and the thermal state of quark matter, respectively.}
\label{table1}
\begin{tabular*}{10cm}{@{\extracolsep{\fill}}c|c|c|c}
  \hline
  $t$ (fm/$c$)             & 0.2    & 0.52   & 0.86  \\
  \hline
  $n$ (${\rm fm}^{-3}$)    & 32.4   & 17     & 10.8  \\
  \hline
  $\bar {\lambda}$ (fm)    & 0.073  & 0.139  & 0.218 \\
  \hline
  $k_{g\bot {\rm min}}$ (GeV/$c$)   & 2.7    & 1.42  & 0.91 \\
  \hline
\end{tabular*}
\end{table}

\end{document}